\pdfoutput=1
\documentclass[acmtog]{acmart}

\setcopyright{cc}
\setcctype{by}
\acmJournal{TOG}
\acmYear{2026}
\acmDOI{10.1145/3859874}

\usepackage{booktabs} %
\usepackage{todonotes, diagbox, multirow}
\definecolor{custompink}{RGB}{237,2,140}
\usepackage{makecell}
\usepackage{float} %

\usepackage[linesnumbered,ruled,vlined]{algorithm2e} %

\SetAlFnt{\small}
\SetAlCapFnt{\small}
\SetAlCapNameFnt{\small}
\SetAlCapHSkip{0pt}

\definecolor{custompink}{RGB}{237,2,140}
\definecolor{darkgreen}{rgb}{0.0, 0.4, 0.0} 
\definecolor{revisionblue}{RGB}{0,114,178}
\definecolor{revisionorange}{RGB}{230,159,0}
\newcommand*{\Representation}{SkinTokens\xspace}

\newcommand{\Framework}{TokenRig\xspace}

\newcommand{\cyp}[1]{\textcolor{black}{#1}}
\newcommand{\added}[1]{#1}
\newcommand{\revised}[1]{#1}
\newcommand{\gain}[1]{_{\,\text{\color{red}(+#1\%)}}}

\usepackage{pifont}

\begin{document}

\title{\Representation: A Learned Compact Representation for Unified Autoregressive Rigging}

\author{Jia-Peng Zhang}
\email{zjp24@mails.tsinghua.edu.cn}
\affiliation{%
 \institution{BNRist, Department of Computer Science and Technology, Tsinghua University}
 \city{Beijing}
 \country{China}}
\author{Cheng-Feng Pu}
\email{pcf22@mails.tsinghua.edu.cn}
\affiliation{%
 \institution{Zhili College, Tsinghua University}
 \city{Beijing}
 \country{China}}
\author{Meng-Hao Guo}
\email{gmh20@mails.tsinghua.edu.cn}
\affiliation{%
 \institution{BNRist, Department of Computer Science and Technology, Tsinghua University}
 \city{Beijing}
 \country{China}}
\author{Yan-Pei Cao}
\email{caoyanpei@gmail.com}
\affiliation{%
\institution{VAST}
\city{Beijing}
\country{China}}
\author{Shi-Min Hu}
\authornote{Corresponding author.}
\email{shimin@tsinghua.edu.cn}
\affiliation{%
\institution{BNRist, Department of Computer Science and Technology, Tsinghua University}
\city{Beijing}
\country{China}}

\begin{abstract}
\cyp{The rapid proliferation of generative 3D models has created a critical bottleneck in animation pipelines: rigging. Existing automated methods are fundamentally limited by their approach to skinning, treating it as an ill-posed, high-dimensional regression task that is inefficient to optimize and is typically decoupled from skeleton generation. We posit this is a representation problem and introduce \textbf{\Representation}: a learned, compact, and discrete representation for skinning weights. By leveraging an FSQ-CVAE to capture the intrinsic sparsity of skinning, we reframe the task from continuous regression to a more tractable token sequence prediction problem. This representation enables \textbf{\Framework}, a unified autoregressive framework that models the entire rig as a single sequence of skeletal parameters and \Representation, learning the complicated dependencies between skeletons and skin deformations. The unified model is then amenable to a reinforcement learning stage, where tailored geometric and semantic rewards improve generalization to complex, out-of-distribution assets. Quantitatively, the \Representation representation leads to a \textbf{98\%–133\%} \revised{relative improvement in skinning-weight $L_1$ error} over state-of-the-art methods, while the full \Framework framework, refined with RL, \revised{improves the skeletal Chamfer distance} by \textbf{17\%–22\%}. Our work presents a unified, generative approach to rigging that yields higher fidelity and robustness, offering a scalable solution to a long-standing challenge in 3D content creation.}
\end{abstract}


\begin{teaserfigure}
  \includegraphics[width=\textwidth]{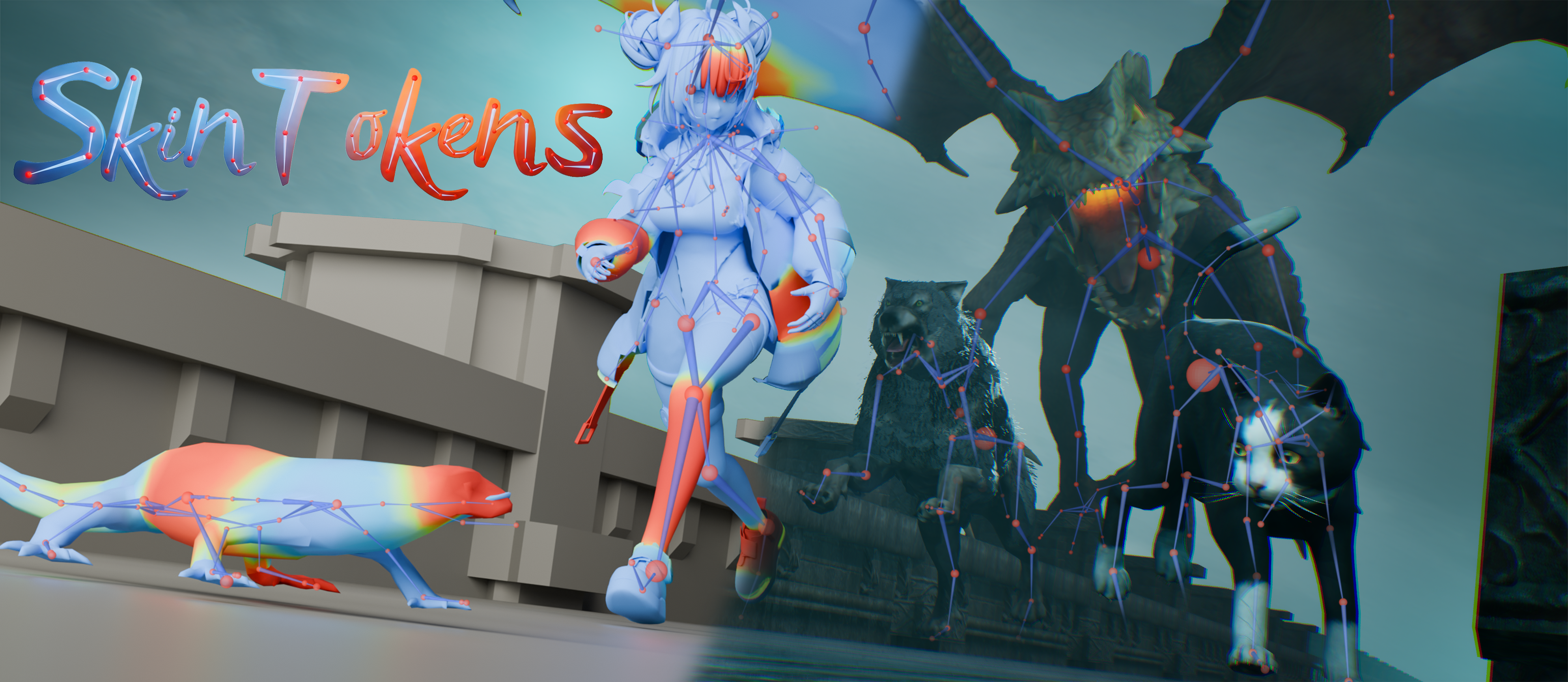}
  \vspace{-4mm}
  \caption{Automated rigging with \Framework. We present \Framework, a unified generative framework that produces high-quality rigs for diverse 3D assets. By leveraging \Representation, i.e., our novel, discrete representation for skinning weights, our method robustly generates high-fidelity skeletons and precise skinning maps (visualized as heatmaps) for complex, real-world geometries, ranging from stylized anime characters to quadrupeds and fantasy creatures. 
  }
  \label{fig:teaser}
\end{teaserfigure}

\maketitle

\section{Introduction}

\cyp{The rapid advancement of generative models has enabled the creation of 3D assets and animations at an unprecedented scale~\cite{xiang2024structured,jiang2024survey,guo2025deep,zhang2024clay,li2025triposg,DBLP:journals/cvm/WangPLGH25,DBLP:journals/cvm/ShiFGZ24}. This progress, however, has exposed a critical and long-standing bottleneck in the digital content pipeline: rigging. And skeleton-driven animation~\revised{\cite{DBLP:conf/siggraph/LewisCF00, kavan2007skinning, magnenat1989joint}} is still an important part of the computer graphics industry, with rigging being an indispensable component. The manual process of creating a skeleton and assigning skinning weights—essential for animation—remains a highly specialized, labor-intensive task. This fundamental mismatch in scalability between asset generation and rigging readiness must be addressed to unlock the full potential of modern 3D AI.}

\cyp{Research in automatic rigging has produced a variety of solutions. Many established approaches rely on template-based skeleton prediction~\revised{\cite{li2021learning, blackman2014rigging,accurig}}, where joint coordinates are fitted to a fixed topology. While effective for common categories like humanoids, these methods exhibit limited generalizability. In contrast, template-free methods\cite{xu2022morig, ma2023tarig} such as RigNet~\cite{xu2020rignet} infer joint likelihoods via heatmaps before constructing skeletal edges, offering greater flexibility. More recently, inspired by the success of large language models, a new class of autoregressive approaches has emerged~\cite{song2025magicarticulate, zhang2025unirig, deng2025anymate, song2025puppeteer, liu2025riganything, guo2025make, guo2025auto}. By serializing the skeletal hierarchy into a sequence of tokens, these models leverage the power of Transformer~\cite{vaswani2017attention} architectures to achieve remarkable generalization in skeleton generation.}

\cyp{Yet, despite this rapid progress in skeletal generation, the equally critical task of skinning weight prediction remains a significant and largely unsolved challenge. The vast majority of prior work treats skinning as a separate, downstream problem, focusing on network architectures that regress the entire skinning matrix from geometric features~\cite{xu2020rignet,liu2019neuroskinning}. This approach is fraught with fundamental issues. \textit{First}, the direct regression of a massive, high-dimensional $N \times J$ matrix \revised{(where $N$ denotes the number of mesh vertices and $J$ the number of skeletal joints)} is an ill-posed and inefficient learning problem. Skinning matrices are intrinsically \textit{sparse}, but dense regression with standard losses like Mean Squared Error (MSE) struggles to enforce this prior, leading to noisy weights that manifest as visually jarring artifacts during motion. \textit{Second}, many methods exhibit an excessive reliance on auxiliary geometric descriptors, such as geodesic distances~\cite{baran2007automatic,dionne2013geodesic}. This renders them fragile in realistic scenarios where meshes may be non-watertight or composed of disconnected components, leading to coarse or unreliable feature estimation.}

\cyp{Most critically, the pervasive decoupling of skeleton and skinning prediction into independent models~\cite{zhang2025unirig} creates a conceptual barrier. This separation prevents any mutual reinforcement between the two tasks; the skeleton is generated without knowledge of the surface deformation it will induce, and the skinning is predicted for a fixed, potentially suboptimal, skeletal structure. This architectural choice inherently constrains the performance ceiling of the entire system and is further compounded by the relative scarcity of datasets with comprehensive skeleton and skinning annotations~\cite{deitke2024objaverse}.}

\cyp{We argue that the path forward requires \textit{a unified model}, and that such unification is only possible with a fundamental shift in the \textit{representation} of skinning weights. \revised{Concretely, our framework casts automatic rigging as a single generative pipeline that, given a raw 3D mesh, proceeds through three stages to produce a complete, animation-ready rig.} \revised{\textbf{Stage~1: a discrete skinning representation.}} To this end, we introduce \textbf{\textit{\Representation}}, a learned, compact, and discrete representation that reframes skinning from a continuous regression problem to a discrete token prediction task. We employ a Finite Scalar Quantized Variational Autoencoder (FSQ-CVAE)~\cite{mentzer2023finite, sohn2015learning}, conditioned on local mesh geometry, to compress the sparse weight assignments for each bone into a short sequence of discrete \Representation.}

\cyp{\revised{\textbf{Stage~2: unified autoregressive generation.}} This representation enables \textit{\textbf{\Framework}}, a unified, end-to-end autoregressive framework. \Framework generates a single, coherent sequence that interleaves skeletal parameters with their corresponding \revised{\Representation}. This holistic formulation allows the model to learn the complex, cross-modal dependencies between skeletal placement and surface skins, a critical relationship ignored by prior decoupled approaches. \revised{Because the entire rig is produced as a single sequence, the Transformer's self-attention conditions the skinning of each bone on all joint positions and bone types, capturing the long-range dependencies that local, per-bone predictors cannot.} \revised{\textbf{Stage~3: reinforcement learning refinement.}} Furthermore, the generative nature of \Framework makes it uniquely suited for refinement using policy gradient methods~\cite{rafailov2023direct, schulman2017proximal, shao2024deepseekmath}. We introduce a reinforcement learning stage with a set of carefully designed reward functions that encode high-level principles of rig quality, such as bone-mesh alignment and deformation smoothness. This allows \Framework to generalize far beyond its training data, successfully rigging complex, ``in-the-wild'' assets where purely supervised methods often fail.}

Our main contributions can be summarized as follows:

\begin{itemize}
\item \cyp{\textbf{A learned discrete representation for skinning weights}, \textit{\textbf{\Representation}}, that transforms skinning from a high-dimensional regression task into a compact sequence prediction problem.}

\item \cyp{\textbf{A unified autoregressive framework}, \textbf{\textit{\Framework}}, that jointly models skeleton generation and skinning, capturing their mutual dependencies for higher-fidelity results.}

\item \cyp{\textbf{A reinforcement learning framework for rig refinement}, with novel reward functions designed to improve the generalization and robustness of the generated rigs on out-of-distribution 3D models.}

\end{itemize}

\section{Related Works}

\subsection{Automatic Rigging Methods}

\subsubsection{Traditional Approaches}
Early research~\cite{baran2007automatic, tagliasacchi2009curve} largely relied on geometric heuristics to infer skeletal structures without data-driven priors. Seminal works like Pinocchio~\cite{baran2007automatic} utilize signed distance fields to approximate the medial surface, refining the embedding via multiplicative optimization. Other topological methods leverage medial representations, such as Voxel Cores~\cite{yan2018voxel} and Erosion Thickness~\cite{yan2016erosion}, to extract skeletons. While these methods function reliably on watertight, geometrically coherent models, they lack semantic understanding and often necessitate significant manual post-processing. Tools like LazyBones~\cite{lazy-bones} for Blender~\cite{blender} automate parts of this process but still fundamentally rely on artists to edit and finalize the rig.

\subsubsection{Learning-Based Approaches}
The advent of deep learning and large-scale datasets, such as ArticulationXL 2.0~\cite{song2025magicarticulate} and Rig-XL~\cite{zhang2025unirig}, has shifted the paradigm toward data-driven methods. Template-based approaches~\cite{li2021learning, ma2023tarig, chu2024humanrig} achieve high fidelity by fitting fixed templates (e.g., humanoids) but fail to generalize to arbitrary characters. Template-free methods like RigNet~\cite{xu2020rignet} and MoRig~\cite{xu2022morig} overcome this by employing Graph Neural Networks (GNNs) to predict joint heatmaps, subsequently connecting them via Minimum Spanning Tree (MST) algorithms. However, MST-based connectivity is sensitive to noisy predictions, often yielding topologically inconsistent results. DRiVE~\cite{sun2024drive} attempts to mitigate this using a diffusion-based framework for robust joint placement.

Most recently, inspired by the sequential editing workflows commonly observed in professional 3D software and the success of Large Language Models (LLMs)~\cite{qwen3, zhang2025bee}, researchers have reformulated skeleton generation as an autoregressive sequence modeling task. These methods~\cite{song2025magicarticulate, zhang2025unirig, deng2025anymate, song2025puppeteer, liu2025riganything, guo2025make, guo2025auto} discretize the skeletal hierarchy into tokens, leveraging Transformers to capture global structural dependencies. Reinforcement learning has also been explored to refine topology, as seen in AutoConnect~\cite{guo2025auto}. Despite this progress, these pipelines predominantly treat skeleton prediction and skinning as decoupled stages. This separation precludes mutual information exchange and necessitates complex feature engineering. In contrast, our method unifies these modalities into a single end-to-end framework, reducing inter-stage error propagation and improving generalization.

\subsection{Automatic Skinning Weight Prediction}

\subsubsection{Traditional Methods}
Classic skinning methods derive weights from geometric properties. Graph-based techniques~\cite{katz2003hierarchical} typically employ min-cut algorithms to segment vertices. Pinocchio~\cite{baran2007automatic} popularized heat diffusion, solving the steady-state heat equation to propagate influence from bone junctions. While prevalent in industrial tools due to their robustness, these methods produce weights purely based on geometry, lacking the semantic nuance required for expressive animation.

\subsubsection{Learning-Based Methods}
Modern methods aim to learn statistical skinning priors from data. RigNet~\cite{xu2020rignet} combines GNNs with geodesic distance features to regress weights directly. Similarly, NeuroSkinning~\cite{liu2019neuroskinning} utilizes graph convolutions with multi-head attention, while Neural Blend Shapes~\cite{li2021learning} employs MeshCNN~\cite{hanocka2019meshcnn} for local feature extraction. Alternative representations have also been proposed; SkinCells~\cite{larionov2025skincells} introduces Voronoi-based sparse weight fields, and MagicArticulate~\cite{song2025magicarticulate} adopts a diffusion formulation to predict residuals relative to geodesic distances. Skinning-inspired architectures have also been applied to cloth deformation for skeleton-driven characters~\cite{DBLP:journals/cvm/LiTYTYLAK24}.

However, a fundamental limitation persists across these methods: they typically frame skinning as a high-dimensional regression problem. Regressing dense matrices for meshes with disconnected components or non-watertight geometry is notoriously unstable and computationally expensive. Furthermore, reliance on auxiliary descriptors like geodesic distance~\cite{zhang2025unirig, xu2020rignet, sun2024drive} limits robustness on complex topologies. Our approach diverges by introducing \Representation, a discrete compression scheme that circumvents direct high-dimensional regression, enabling scalable and robust joint learning of skinning and structure.

\added{
\section{Preliminary}
\label{sec:preliminary}

\subsection{Vector Quantized Variational Autoencoder}
The Vector Quantized Variational Autoencoder (VQ-VAE)~\cite{DBLP:conf/nips/OordVK17} is a foundational probabilistic generative model that simultaneously performs data compression and representation learning by mapping continuous latent variables into a discrete codebook. It discretizes the encoder output \(z \in \mathbb{R}^d\) by replacing each vector with the nearest entry from a learned codebook \(\mathcal{C} = \{c_1, \ldots, c_K\}\):
\[
\hat{z} = \arg\min_{c_i \in \mathcal{C}} \|z - c_i\|_2^2.
\]
Because the argmin quantization operation is non-differentiable, VQ-VAE approximates gradients using the Straight-Through Estimator (STE)~\cite{bengio2013estimating}. However, training VQ-VAE is notoriously difficult: it requires careful balancing of auxiliary losses (e.g., commitment and codebook losses) and relies on heuristics such as exponential-moving-average (EMA) updates or codebook re-seeding~\cite{DBLP:conf/nips/RazaviOV19} to avoid codebook collapse, where only a fraction of the codebook is utilized.

\subsection{Finite Scalar Quantization}
Finite Scalar Quantization (FSQ)~\cite{mentzer2023finite} simplifies the vector quantization process by replacing the learnable codebook with a fixed, implicitly defined grid. Instead of maintaining a separate dictionary, FSQ bounds each encoded latent dimension via a nonlinear mapping (e.g., sigmoid or tanh) and subsequently rounds it to the nearest discrete level:
\[
\hat{z} = \operatorname{Round}\!\big((L-1)\,\sigma(z)\big), \quad 
\hat{z} \in \{0,1,\dots,L-1\}^d,
\]
yielding an implicit codebook of size \(|\mathcal{C}| = L^d\). 
Similar to VQ, gradients are propagated using the STE approximation:
\[
\operatorname{Round}_{\text{ste}}(x) = x + \text{sg}(\operatorname{Round}(x) - x),
\]
where $\text{sg}(\cdot)$ denotes the stop-gradient operator. 

Crucially, compared to naive VQ-VAE, FSQ offers two key advantages: (1) \textbf{No codebook collapse}: the fixed grid naturally encourages full utilization of the latent space, avoiding dead codes; and (2) \textbf{Simple training}: it removes the auxiliary commitment and codebook losses together with the EMA or re-seeding heuristics. FSQ thus serves as a robust, drop-in replacement for VQ at substantially reduced complexity.
}

\section{Method}

\subsection{\cyp{Overview}}
\begin{figure*}[h!]
    \centering
    \includegraphics[width=\linewidth]{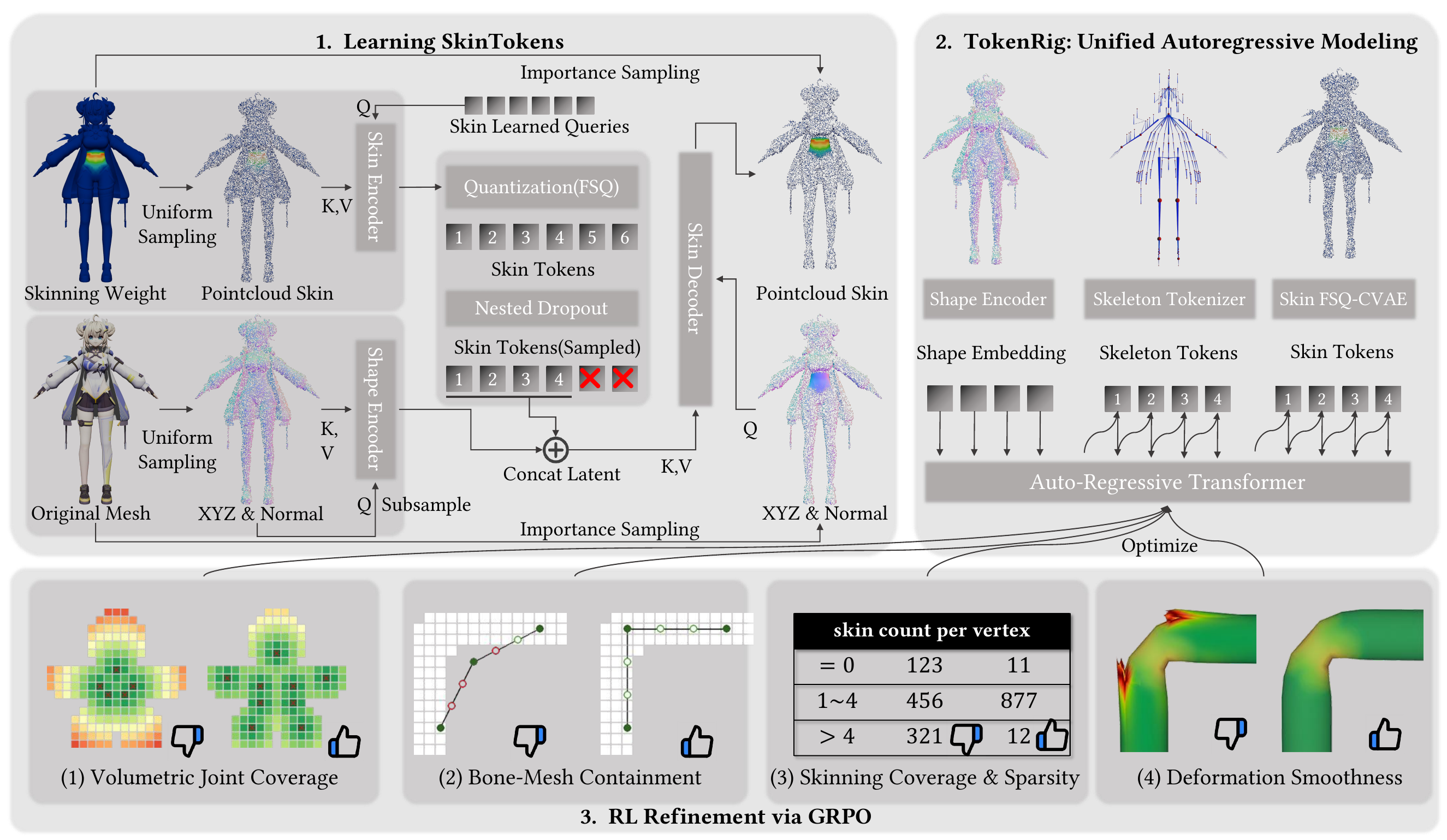}

    \caption{\textbf{Overview of the \Framework Framework.} Our method consists of three key stages: \textbf{(1) Learning \Representation (Section~\ref{sec:FSQ-CVAE}):} We first train a FSQ-CVAE~\cite{kingma2013auto,sohn2015learning,mentzer2023finite} to compress sparse skinning weights into a compact, discrete representation. Mesh geometry and skinning weights are processed by VecSet~\cite{zhang20233dshape2vecset} encoders, and the resulting features are discretized into \Representation via Finite Scalar Quantization (FSQ)~\cite{mentzer2023finite}. We employ nested dropout~\cite{bachmann2025flextok,rippel2014learning} and importance sampling to ensure robust reconstruction of active deformation regions.
    \textbf{(2) Unified Autoregressive Modeling (Section~\ref{sec:method-generation}):} We formulate rigging as a sequence generation task. A Transformer generates a single, unified sequence comprising the complete skeleton followed by the learned \Representation (from Stage 1), conditioned on global shape embeddings to capture structural dependencies. \textbf{(3) RL Refinement via GRPO (Section~\ref{sec:GRPO}):} To improve generalization to in-the-wild assets, we fine-tune the model using Group Relative Policy Optimization (GRPO)~\cite{liu2024deepseek}. We introduce four specific rewards: Volumetric Joint Coverage (ensuring bone distribution), Bone-Mesh Containment (preventing protrusion), Skinning Coverage and Sparsity (ensuring valid weighting), and Deformation Smoothness (preventing artifacts during animation).}
    
    \label{fig:FSQ-CVAE}
\end{figure*}

\cyp{Our method reframes automatic rigging as a unified, generative sequence modeling task \textbf{(see Figure~\ref{fig:FSQ-CVAE})}. This is achieved through three key stages. First, we introduce \textit{\Representation}, a novel discrete representation for skinning weights learned via a FSQ-CVAE~\cite{mentzer2023finite, sohn2015learning}. This representation transforms the intractable problem of regressing a high-dimensional, sparse matrix into a tractable token prediction task (Section~\ref{sec:method-representation}). Second, this representation enables \textit{\Framework}, a unified autoregressive Transformer that learns to generate a single, interleaved sequence of skeletal parameters and their corresponding \Representation, thereby jointly modeling the entire rig (Section~\ref{sec:method-generation}). Finally, we employ a reinforcement learning fine-tuning stage using a set of tailored reward functions, which significantly enhances the model's generalization capabilities to complex, out-of-distribution 3D assets (Section~\ref{sec:GRPO}).}

\subsection{\cyp{\Representation: A Learned Discrete Representation for Skinning}}
\label{sec:method-representation}

\cyp{The core of our framework is a novel representation for skinning weights that circumvents the fundamental limitations of direct regression. We motivate and detail this representation below.}

\subsubsection{\cyp{The Sparsity and Challenge of Skinning}}

\cyp{Formally, given a mesh $\mathcal{M} = \{\mathcal{V} \in \mathbb{R}^{3 \times N}, \mathcal{F}\}$ with $N$ vertices and a skeleton with $J$ joints, the skinning task is to predict the $N \times J$ weight matrix $\mathcal{W}$. In production models, $N$ can exceed $10^5$ and $J$ can exceed $10^2$, leading to matrices with over $10^7$ elements. Directly regressing such a large matrix is computationally demanding and statistically challenging.}

\cyp{Crucially, the skinning matrix $\mathcal{W}$ is intrinsically sparse. Practically, each vertex is typically influenced by no more than four joints, meaning the number of non-zero elements is at most $4N$. As shown in Table~\ref{tab:sparsity_dataset}, the average sparsity ratio across several public datasets is extremely low ($2-10\%$). This severe class imbalance makes training with standard dense losses like Mean Squared Error (MSE) highly inefficient, \revised{as the overwhelming majority of zero-valued entries dominate the loss, causing the model to under-prioritize the sparse but critical non-zero weights}. Furthermore, the arbitrary ordering of vertices in a mesh makes it impossible to apply traditional sparse matrix compression techniques that rely on structured sparsity. These challenges motivate a learned approach to compression.}

\begin{table}[t]
    \centering
    \resizebox{\columnwidth}{!}{  %
    \begin{tabular}{l|c|c|c}
        \hline
        & \makecell{ModelsResource} & \makecell{VRoid Hub} & \makecell{Articulation 2.0} \\
        \hline
        avg. $N$ & 1297.69 & 16929.78 & 6247.05 \\
        avg. $J$ & 19.87 & 95.56 & 34.46 \\
        avg. $\sum \mathbb{I} [w>0]$ & 1700.18 & 28392.74 & 12655.45 \\
        avg. sparsity & 7.40\% & 2.43\% & 9.38\% \\
        \hline
    \end{tabular}
    }
    \caption{\textbf{Sparsity Analysis of Skinning Weights.} Statistics across three major datasets~\cite{Models-Resource,isozaki2021vroid,song2025magicarticulate} reveal that skinning matrices are extremely sparse ($<10\%$ active weights), motivating our design of the compressed \Representation representation.}
    \label{tab:sparsity_dataset}
\end{table}

\subsubsection{\cyp{FSQ-CVAE for Skinning}\label{sec:FSQ-CVAE}}

\cyp{To address these challenges, we propose compressing the skinning weights for each individual bone $j \in [1, J]$, denoted as $\mathcal{W}^* = \{w_{(\cdot), j}\}$ 
, into a compact, discrete representation. Our approach is based on the Conditional Variational Autoencoder (CVAE) framework~\cite{kingma2013auto,sohn2015learning}, which learns a latent representation of data $x$ conditioned on an observed variable $y$. Here, the skinning weights $\mathcal{W}^*$ are the data to be reconstructed, conditioned on the full mesh geometry $\mathcal{M}$.}

\cyp{As illustrated in Figure~\ref{fig:FSQ-CVAE}, our architecture consists of two distinct encoders, $E_M$ and $E_W$, which follow the VecSet design~\cite{zhang20233dshape2vecset} to process the mesh and weights as unordered point sets. The mesh encoder $E_M(\mathcal{M})$ produces shape features, while the skin encoder $E_W(\mathcal{W}^*)$ produces latent weight features $L_W$. The objective is to convert this continuous latent representation into a discrete format suitable for sequence modeling. We refer to this learned, quantized representation as \Representation.
This crucial discretization step is performed by applying Finite Scalar Quantization (FSQ)~\cite{mentzer2023finite} to the latent features $L_W$. Unlike traditional vector quantization~\cite{van2017neural}, FSQ avoids a learnable codebook by quantizing each latent dimension to the nearest level on a fixed grid. This simplifies training, requires no auxiliary losses, and provides excellent codebook utilization. Gradients are passed through the non-differentiable quantization step using the Straight-Through Estimator (STE)~\cite{bengio2013estimating}.}

\cyp{The resulting discrete tokens $L_D = \text{FSQ}(L_W)$ are then concatenated with the shape features from $E_M$ and passed to a decoder. To improve robustness and encourage a more compositional representation, we adopt a nested dropout scheme analogous to FlexTok~\cite{bachmann2025flextok,rippel2014learning}, randomly selecting a prefix of the token sequence during training. The decoder reconstructs the per-vertex skinning weights $\mathcal{W}^*_{\text{pred}}$, with a final sigmoid activation to ensure outputs are bounded in $[0,1]$.}

\subsubsection{\cyp{Loss Function for Sparse Weight Reconstruction}}\label{sec:loss}

Standard VAEs often assume a Gaussian likelihood~\cite{kingma2013auto}, optimized with Mean Squared Error (MSE). However, as our outputs represent probability-like skinning weights in the range $[0, 1]$, a Bernoulli likelihood with a Binary Cross-Entropy (BCE) loss is more suitable.

\begin{figure}
    \centering
    \includegraphics[width=1\linewidth]{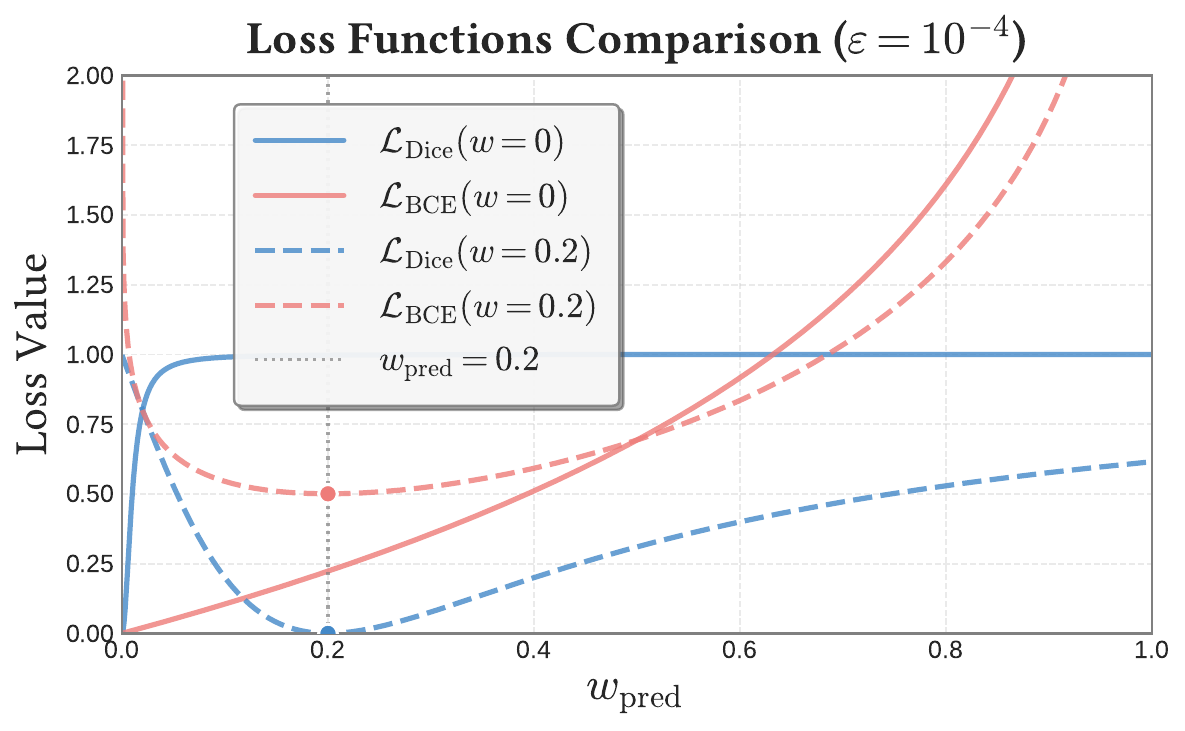}
    \caption{\textbf{Gradient Analysis of Loss Functions.} A comparison of Binary Cross Entropy (BCE) and Dice loss~\cite{sudre2017generalised} landscapes for a target weight $w=0.2$. While both minimize at the correct value, Dice loss provides significantly larger gradients for non-zero targets ($w_{\text{pred}} \in [0,1]$), effectively counteracting the extreme sparsity of skinning matrices where BCE gradients tend to vanish.}
    \label{fig:functions}
\end{figure}

\cyp{More importantly, the extreme sparsity of $\mathcal{W}^*$ presents a significant class imbalance. To counteract this, we incorporate the Dice loss~\cite{milletari2016v}, a metric widely used in image segmentation that is equivalent to the F1-score and focuses supervision on positive (non-zero) samples:}
$$
\mathcal{L}_{\text{Dice}} = 1 - \frac{2|\mathcal{W}^*_{\text{pred}} \cap \mathcal{W}^*|}{|\mathcal{W}^*_{\text{pred}}| + |\mathcal{W}^*|} = \sum_{j \le J} 1 - \frac{2 \sum_i {w_{\text{pred}}}_{i,j} w_{i,j} + \varepsilon}{\sum_i {w_{\text{pred}}}_{i,j}^2 + \sum_i w_{i,j}^2 + \varepsilon},
$$
where a small constant $\varepsilon = 10^{-4}$ ensures numerical stability. As shown in Figure~\ref{fig:functions}, the Dice loss provides a stronger gradient signal for non-zero weights compared to BCE, effectively amplifying supervision where it matters most. More specifically, when $w = 0$, the gradient magnitude $\frac{\partial \mathcal{L}}{\partial w_{\text{pred}}}$ is small for $w_{\text{pred}} \in (\varepsilon, 1]$, whereas for $w > 0$, it is considerably larger across $w_{\text{pred}} \in [0,1]$. This selectively amplifies gradients for positive entries, aligning well with the imbalanced sparsity in $\mathcal{W}$. The minimum stationary condition $w_{\min} = w$ holds for $\mathcal{L}_{\text{Dice}}$, $\mathcal{L}_{\text{BCE}}$, and $\mathcal{L}_{\text{MSE}}$, with $\mathcal{L}_{\text{Dice}}(w_{\min}) = 0$, ensuring smooth blending across joint seams. Overall, the final reconstruction loss for the CVAE is a weighted sum of these objectives:
\begin{align*}
\mathcal{L}_{\text{VAE}} = \lambda_{\text{BCE}}\mathcal{L}_{\text{BCE}}(\mathcal{W}^*_{\text{pred}}, \mathcal{W}^*) + \lambda_{\text{MSE}}\mathcal{L}_{\text{MSE}}(\mathcal{W}^*_{\text{pred}}, \mathcal{W}^*) + 
\\
\lambda_{\text{Dice}}\mathcal{L}_{\text{Dice}}(\mathcal{W}^*_{\text{pred}}, \mathcal{W}^*),
\end{align*}
\cyp{where we also retain a small MSE term in our implementation for stability. This composite loss ensures both accurate reconstruction of values and precise localization of sparse, non-zero weight regions.}

\subsubsection{\cyp{Importance Sampling for Efficient Training}}

To accelerate training and focus the model's capacity on critical deformation regions, we employ a hybrid sampling strategy for the mesh points fed to the skin decoder. During each training step, we provide the decoder with a combination of points sampled uniformly from the mesh surface, $\mathcal{P}_{\text{uniform}}$, and points sampled densely from regions with non-zero ground-truth skinning weights, $\mathcal{P}_{\text{dense}}$. While the shape encoder $E_M$ only sees $\mathcal{P}_{\text{uniform}}$ (to match inference conditions), this importance sampling for the decoder ensures that sparse, active deformation zones are well-represented in every training batch, leading to faster convergence and higher fidelity.

\subsection{\cyp{\Framework: Unified Autoregressive Modeling} \label{sec:method-generation}}

\cyp{The discrete and compact nature of \Representation enables us to move beyond the limitations of decoupled, multi-stage pipelines~\cite{song2025magicarticulate, zhang2025unirig, deng2025anymate, song2025puppeteer, liu2025riganything, guo2025make, guo2025auto}. We can now represent the entire rig, i.e., both skeleton and skinning, as a single, coherent sequence of discrete tokens. This allows us to formulate rigging as a unified sequence generation task, which we solve with \Framework, an autoregressive Transformer model that generates the complete skeleton first, followed by the corresponding skinning weights.}

\subsubsection{\cyp{Unified Sequence Representation}}

The power of our approach stems from a novel, coherent sequence representation that jointly captures skeletal structure and surface skins.

\noindent\textbf{Skeletal Tokenization.} Following the methodology of recent work in skeleton generation~\cite{zhang2025unirig}, we first serialize the skeletal hierarchy. Joint coordinates $\mathcal{J}$ are uniformly quantized and represented as a sequence of discrete integer tokens $d_i = (dx_i, dy_i, dz_i)$. The bone order is established using predefined templates (e.g., for bipeds) or chain partitioning strategies for more general structures:
\[
\begin{aligned}
\textbf{<bos>}~\textbf{<type}_1\textbf{>}~dx_1~dy_1~dz_1~dx_2~dy_2~dz_2 \cdots \textbf{<type}_2\textbf{>} \dots \\
\textbf{<type}_k\textbf{>}~dx_t~dy_t~dz_t \dots dx_T~dy_T~dz_T~\textbf{<eos>}.
\end{aligned}
\]
Here, \textbf{<bos>} and \textbf{<eos>} denote sequence boundaries, and each bone chain is prefixed with a special \textbf{<type>} token that serves as a categorical identifier (e.g., \textit{mixamo}).

\noindent\cyp{\textbf{Sequential Composition of \Representation.} Following the complete skeletal sequence, we introduce the skinning information as a subsequent sequence of \Representation. For each bone $i$ in the skeleton, its corresponding skinning influence is represented by a sequence of $\mathcal{T}_D$ discrete \Representation from our pre-trained FSQ-CVAE (see Sec.~\ref{sec:FSQ-CVAE}). These individual \Representation sequences are then concatenated in canonical order to form a single, continuous block representing the skinning for the entire model.
The complete autoregressive sequence is thus a structured composition of these two modalities:
\[
\begin{aligned}
\textbf{<bos>}~\textbf{<type}_1\textbf{>}~dx_1~dy_1~dz_1 \cdots \textbf{<type}_k\textbf{>}~dx_T~dy_T~dz_T \\
{\mathcal{D}}_{1,0} \dots \mathcal{D}_{1,\mathcal{T}_D} \dots \mathcal{D}_{T,0} \dots \mathcal{D}_{T,\mathcal{T}_D}~\textbf{<eos>},
\end{aligned}
\]
where $\mathcal{D}_{i,j}$ denotes the $j$-th token of the $i$-th joint’s skin latent.}

\cyp{This sequential, two-part structure allows the generation of \Representation to be globally conditioned on the fully generated skeleton. The Transformer's self-attention mechanism can access all joint positions and bone types when predicting the skinning for any given bone, enabling it to model complex, long-range dependencies. This holistic conditioning is a significant advantage over methods that predict skinning based only on local features.}

\subsection{\cyp{Generalization via Reinforcement Learning Refinement}}\label{sec:GRPO}

\cyp{While the supervised \Framework model captures the statistical distribution of rigs in the training data, it faces inherent limitations when applied to out-of-distribution (OOD) assets. Due to the nature of next-token prediction, the model may default to ``average'' solutions or fail to capture global geometric constraints on complex topologies (e.g., missing auxiliary limbs like wings or tails, or placing bones outside the mesh). To address this, we introduce a post-training refinement stage using Reinforcement Learning (RL).}

\cyp{Unlike prior work that relies on costly annotated preference data (DPO)~\cite{guo2025auto,rafailov2023direct}, we leverage Group Relative Policy Optimization (GRPO)~\cite{shao2024deepseekmath}. This allows us to optimize the model directly against a set of explicit, non-differentiable geometric and semantic rewards that encode professional rigging criteria.}

\subsubsection{\cyp{Task-Specific Reward Design}}
\cyp{We design a suite of four rewards to guide the model toward topologically valid and functionally robust rigs. These rewards punish common failure modes such as bone protrusion, unconnected vertices, and varying skinning density.}

\paragraph{{(1) \cyp{Volumetric Joint Coverage ($R_{vj}$).}}}
To ensure the generated skeleton adequately spans the geometry of the character, we verify that joints are distributed throughout the mesh's interior volume. We voxelize the mesh into a grid of resolution $r^3$ (where $r=196$). For each voxel center $v_i$, we compute its Euclidean distance to every joint $J_j$, and sum up the results using an exponential kernel. This gives the voxel-joint distance reward:
$$R_{vj}=\frac{1}{V}\sum_{i=1}^{V}\exp{(-\alpha\min_{j=1}^{J}{||v_i-J_j||_{2}})}$$
where $V$ is the number of occupied voxels and $\alpha=0.05$ is a scaling factor that controls the falloff. Intuitively, 
\cyp{this reward encourages the placement of joints in all significant mesh parts, preventing ``missing'' bones in extremities.}

\paragraph{{\cyp{(2) Bone-Mesh Containment ($R_{vk}$).}}}
\cyp{A fundamental rule of rigging is that bones should reside within the character's body. We penalize structural hallucinations where bones protrude outside the mesh surface. We sample $s$ points uniformly along each of the $J$ generated bones and check their containment within the voxelized mesh:}
$$R_{vk}=\frac{1}{J\times (s+1)}\sum_{j=1}^{J}\sum_{i=1}^{s+1}\mathbb{I}[J_{j,i} \in \mathcal{V}]$$
where 
$J_{j,i}$ the $i$-th uniformly sampled point along the $j$-th bone, and 
\cyp{$\mathbb{I}[\cdot]$ is the}
indicator function that returns $1$ if point $J_{j,i}$ lies inside an occupied voxel and $0$ otherwise. This reward directly penalizes bones that protrude outside the mesh, ensuring that the skeleton remains geometrically consistent with the mesh.

\paragraph{{(3) Skinning Coverage and Sparsity ($R_{sc}$).}}
\cyp{Naively optimizing for joint placement can lead to degenerate solutions (e.g., filling the volume with excessive joints). We therefore introduce a skinning reward to enforce two constraints: every vertex must be influenced by at least one bone (\textit{avoiding ``unbound'' geometry}), and no vertex should be influenced by too many bones (\textit{enforcing sparsity}).}
\begin{align*}
R_{sc} &= 1 - \frac{1}{2}R_z - \frac{1}{2}R_m, \\
R_z &= \left(\frac{1}{|\mathcal{V}|}\sum_{i}{\prod_{j=1}^{J}\mathbb{I}[\mathcal{W}_{i,j}<\beta]}\right)^{\alpha_z}, \\
R_m &= \left(\frac{1}{|\mathcal{V}|}\sum_{i}{\mathbb{I}\left[\left(\sum_{j=1}^{J}\mathbb{I}\left[\mathcal{W}_{i,j}>\beta\right]\right)>4\right]}\right)^{\alpha_m}.
\end{align*}
\cyp{Here, $R_z$ measures the fraction of vertices with zero skinning weights (where $\mathcal{W}_{i,j}<\beta$), and $R_m$ measures the fraction of vertices influenced by more than $4$ bones (where $\sum_{j=1}^{J}\mathbb{I}[\mathcal{W}_{i,j}>\beta])>4$), with threshold $\beta=0.1$. $\alpha_z,\alpha_m$ are hyperparameters controlling the penalty strength. This term effectively regularizes the interaction between skeleton and skinning.}

\paragraph{{\cyp{(4) Deformation Smoothness ($R_{mo}$).}}}
Since the ultimate goal of rigging is animation, we assess the quality of deformations under animation. We define a motion reward that penalizes spiky or distorted artifacts. We apply the Linear Blend Skinning (LBS) algorithm to the mesh using randomly sampled poses and measure the distortion of mesh edges:
$$R_{mo}=\left(1+s\cdot\mathbb{E}_{p\sim\mathcal{P}}\left[\max_{e\in\mathcal{E}}{\left(1,\frac{{l}(\text{LBS}(e))}{{l}(e)+\varepsilon}\right)}\right]\right)^{-1}$$
where $l(e)$ is the $\mathcal{L}^2$ length of edge $e$ in the rest pose, $\text{LBS}$ is the linear blend skinning function that deforms the edge under pose $p$, $\mathcal{P}$ represents the possible pose space of the skeleton, $s$ the hyperparameter to scale values and $\varepsilon=10^{-6}$ ensures numerical stability. \cyp{This encourages the model to predict skinning weights that preserve local surface geometry during articulation.} In practice, we approximate the expectation with $5$ randomly sampled poses.

\subsubsection{\cyp{Policy Optimization with GRPO}}

The final reward $R$ is a weighted sum of the above components:
$R = w_{vj}\cdot R_{vj} + w_{vk}\cdot R_{vk} + w_{sc}\cdot R_{sc} + w_{mo}\cdot R_{mo}$. Crucially, if the generated token sequence is invalid (cannot be decoded into a usable rig), we assign $R=0$. 

We refine the policy $\pi_{\theta}$ using GRPO~\cite{shao2024deepseekmath}. For each input mesh, we sample a group of $G$ outputs (i.e., token sequences) $o_i,i\in\{0,1,\cdots,B-1\}$ from the current policy $\pi_{old}$. We verify the structural consistency of each output, decode it to compute the rewards, and calculate the advantage $R_{i}$ for each sample by normalizing the rewards within the group. The training objective is:
\begin{align*}
\mathcal{L}&=\frac{1}{G}\sum_{i=1}^G\frac{1}{|o_i|}\sum_{t=1}^{|o_i|}\min\left[\frac{\pi_\theta(o_{i,t})}{\pi_{old}(o_{i,t})},\text{clip}\left(\frac{\pi_\theta(o_{i,t})}{\pi_{old}(o_{i,t})},1-\epsilon,1+\epsilon\right)\right]R_{i}\\ & - \beta\mathbb{D}_{\text{KL}}[\pi_\theta||\pi_{ref}]
\end{align*}
where $\mathbb{D}_{\text{KL}}$ is the KL divergence~\cite{kl-divergence}. This approach stabilizes training by using group-relative baselines rather than a separate critic network, allowing \Framework to effectively ``self-correct'' its generation logic based on geometric validity.

\begin{figure}[t]
    \centering
    \includegraphics[width=1.0\linewidth]{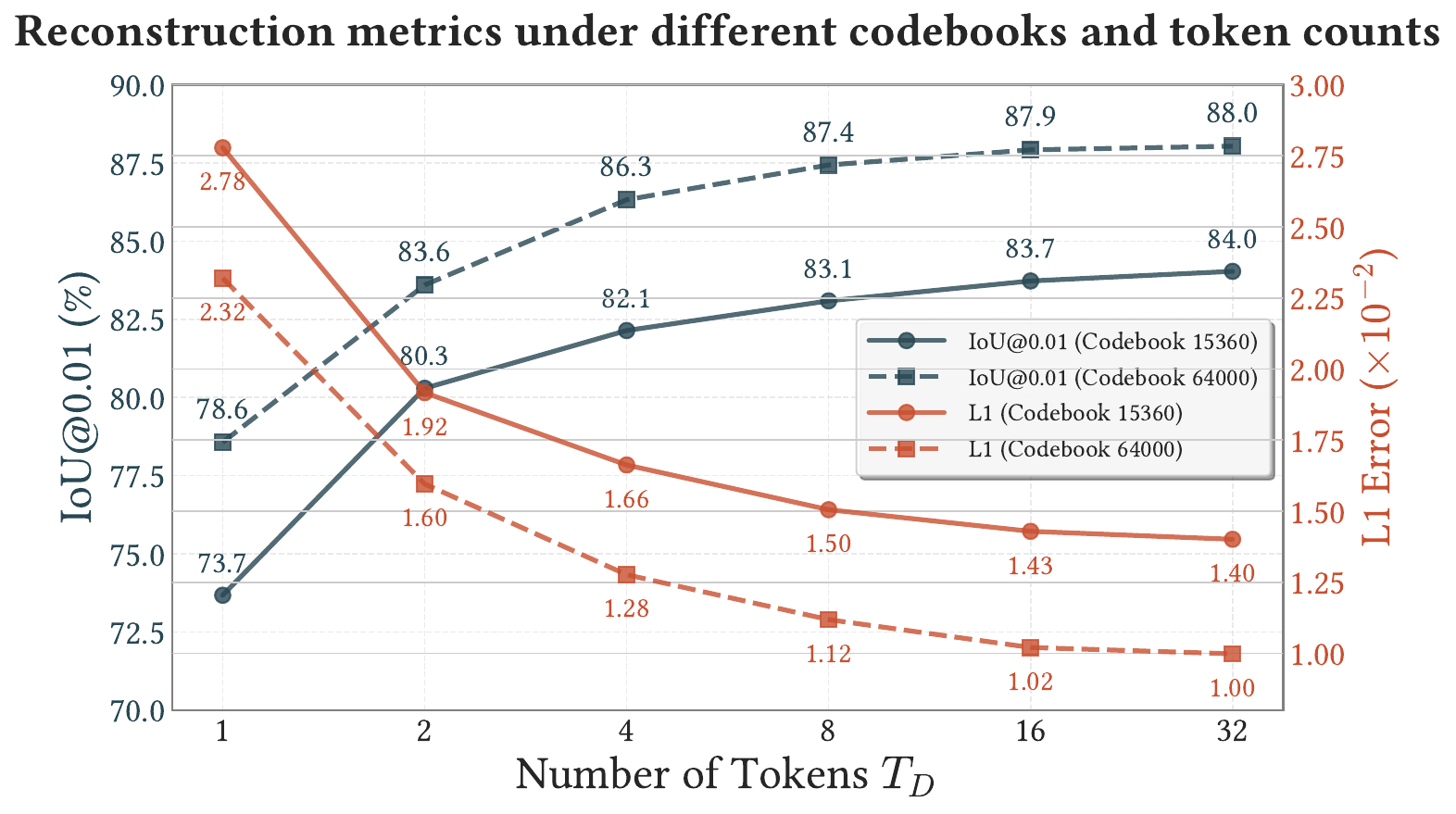}
    \caption{\textbf{\Representation Reconstruction Fidelity.} We evaluate the reconstruction quality (IoU and L1 Error) of the FSQ-CVAE across varying codebook sizes $C$ and token sequence lengths $T_D$. The results demonstrate that \Representation achieve high fidelity with as few as $4$ tokens, validating the compressibility of skinning data. The configuration $C = [8,8,8,6,5] = 15,360$ (lines with circles) is selected for our final model for its superior balance of compression and accuracy. The figure reports the IoU scores at $\varepsilon = 10^{-2}$ and corresponding $L_1$ reconstruction errors on the {Articulation 2.0}~\cite{song2025magicarticulate} test dataset}
    \label{fig:recon_vae}
\end{figure}

\section{Experiment}

\begin{table}[t]
\centering
\resizebox{\columnwidth}{!}{  %
\begin{tabular}{lccc}
\hline
Codebook Size & Total Entries & Utilization & $\text{Compression Ratio}^*$ \\
\hline
$[8, 8, 8, 5, 6]$ & $15360$ & \revised{$99.7\%$} & $208.23\times$ \\
$[8, 8, 8, 8, 8]$ & $32768$ & \revised{$98.5\%$} & $195.22\times$ \\
$[8, 8, 8, 5, 5, 5]$ & $64000$ & \revised{$97.3\%$} & $183.74\times$ \\
\hline
\end{tabular}
}
\caption{\textbf{Codebook Utilization and Compression Analysis.} We compare different FSQ codebook configurations ($C$). $^*$ The compression ratio is computed against a raw FP16 baseline using the {Articulation 2.0}~\cite{song2025magicarticulate} dataset. For example, storing skinning weights for an average mesh ($6,247$ vertices) typically requires $12,494$ bytes; our representation ($T_D=32$ tokens) reduces this to just $64$ ($32\times2$) bytes, achieving a substantial reduction in storage requirements.}
\label{tab:codebook}
\end{table}

\begin{figure}[t]
    \centering
    \includegraphics[width=1.0\linewidth]{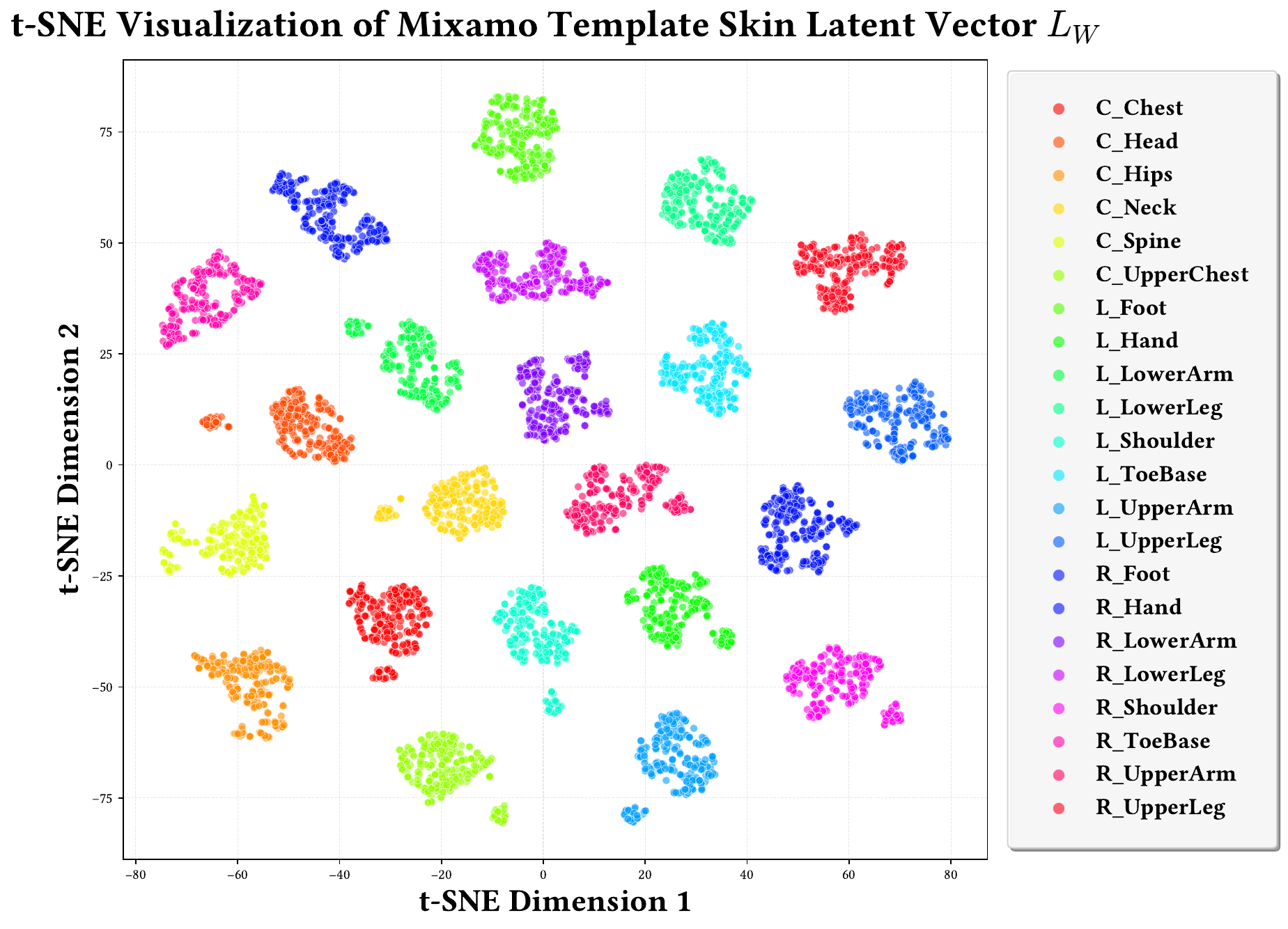}
    \caption{\textbf{Learned Semantics of \Representation.} A t-SNE visualization of the continuous latent vectors $L_W$ prior to quantization, sampled from $300$ instances in the VRoid dataset~\cite{isozaki2021vroid}. Points are colored by bone category (e.g., Head, Hips). The clear emergence of anatomical clusters indicates that the encoder captures a semantic structural prior, learning to represent ``body part concepts'' invariant to specific mesh geometries.}
    \label{fig:t-sne}
\end{figure}

\begin{figure*}[t]
    \centering
    \includegraphics[width=\textwidth]{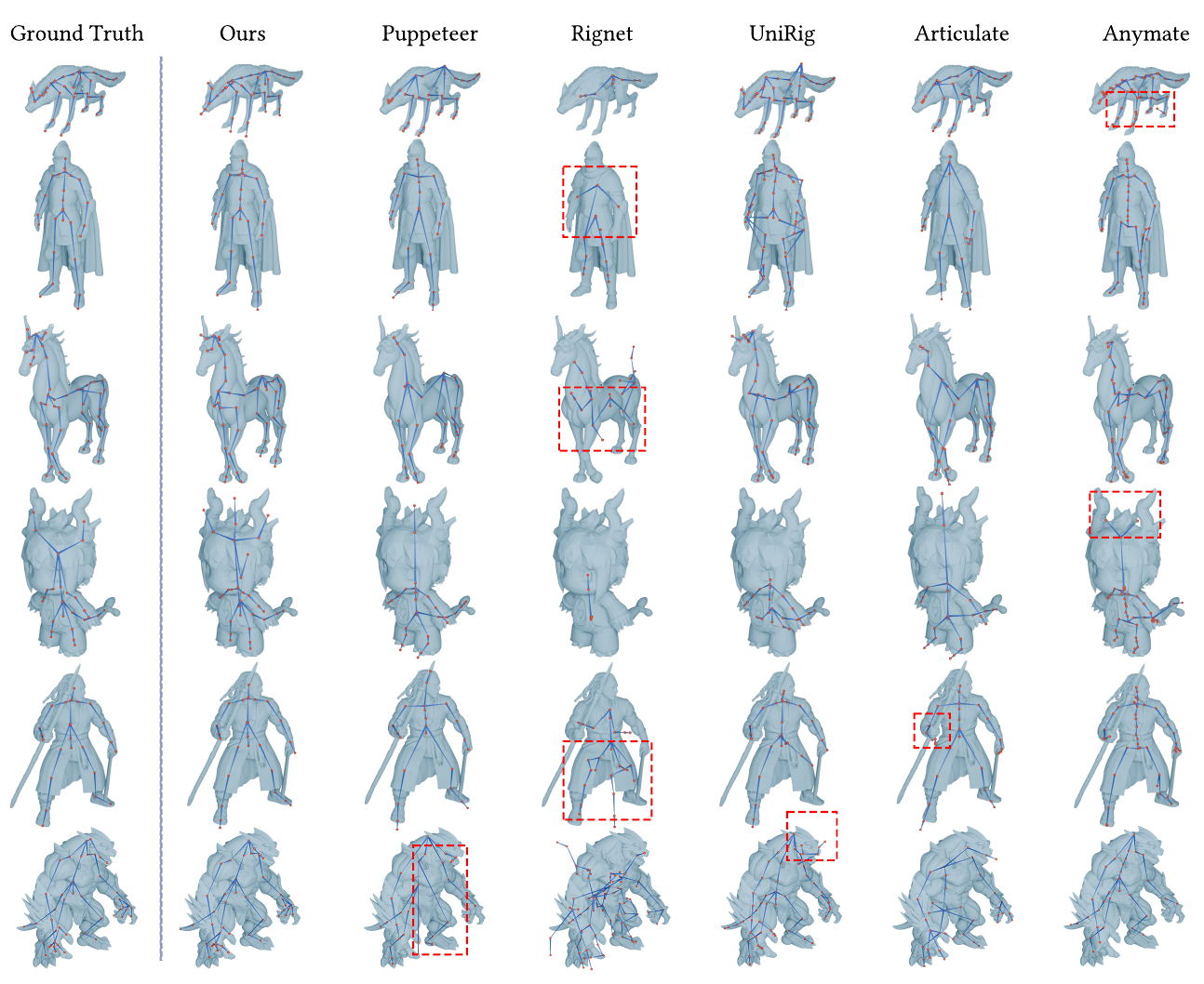}
    \caption{\textbf{Qualitative Comparison of Skeleton Generation.} We compare \Framework (Ours) against state-of-the-art baselines. 
    While baseline methods exhibit partial structures, missing details, or redundant joints, our method synthesizes structurally coherent and semantically faithful skeletons across diverse character types.}
    \label{fig:compare_rigging}
\end{figure*}

\begin{figure*}[t]
    \centering
    \includegraphics[width=\textwidth]{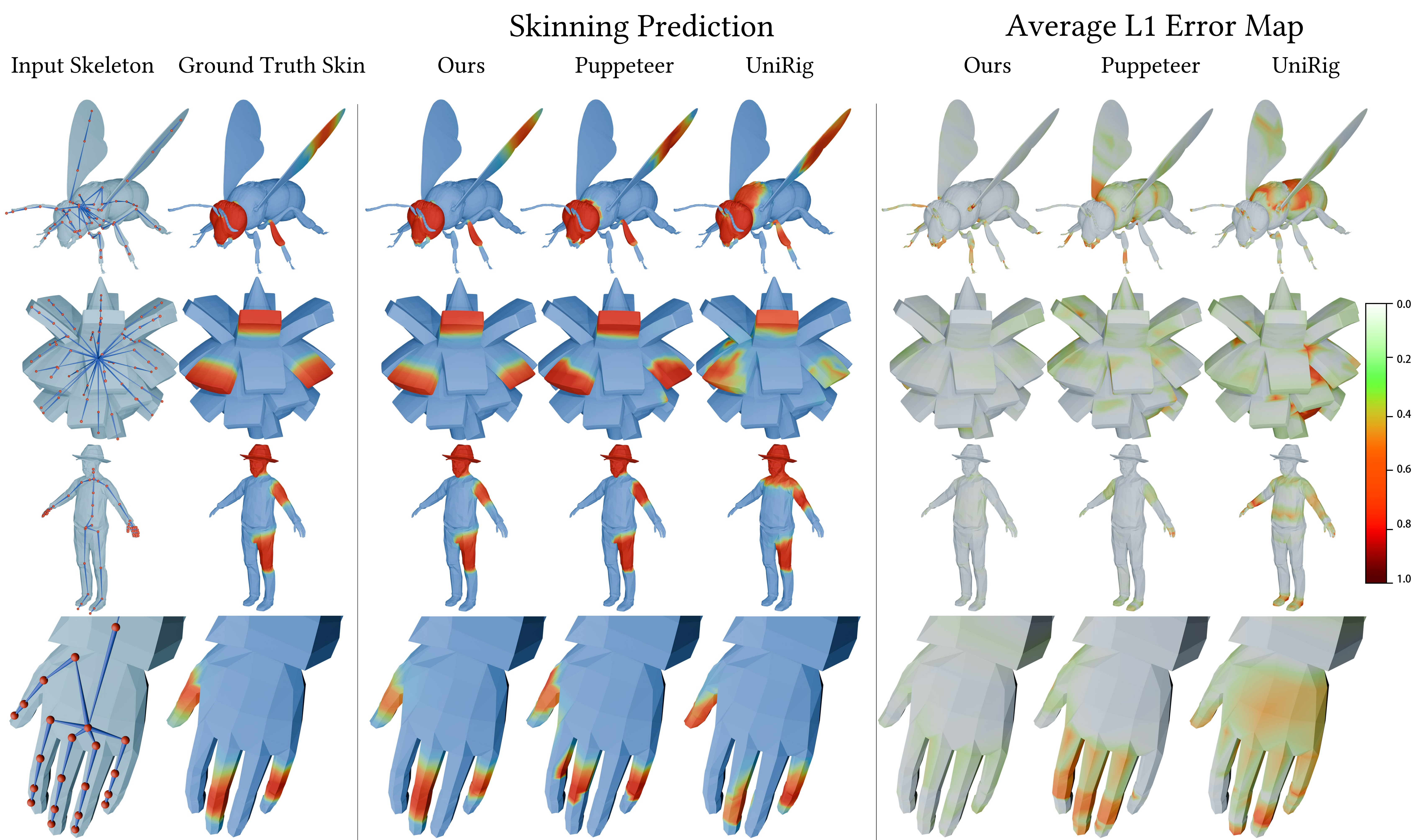}
    \caption{\textbf{Qualitative Comparison of Skinning Prediction.} We visualize predicted skinning weights and the corresponding average L1 error maps. Baseline methods often suffer from ``bleeding'' artifacts, where weights spill onto unconnected mesh parts (see UniRig/Puppeteer columns). \Framework (Ours) produces clean, locally coherent influence maps that closely match the Ground Truth, particularly in fine-grained regions like fingers.}
    \label{fig:compare_skin}
\end{figure*}

\subsection{\cyp{Implementation and Experimental Setup}}

\subsubsection{Dataset Configuration}
\cyp{To ensure our model generalizes across diverse topologies and articulation styles, we construct a composite dataset sourcing from Articulation2.0~\cite{song2025magicarticulate} (70\%), VRoid Hub~\cite{isozaki2021vroid} (20\%), and ModelsResource~\cite{Models-Resource,xu2020rignet} (10\%). This mix balances high-quality, professional rigs with varied community-created assets. All geometry is normalized to the canonical unit cube $[-1,1]^3$.}

\subsubsection{\cyp{Robustness-Oriented Data Augmentation}}
\cyp{We implement a careful augmentation pipeline designed to simulate the topological imperfections and irregular structures found in ``in-the-wild'' assets.}

\begin{enumerate}
\item \cyp{\textbf{Structural Variation:} To prevent overfitting to specific skeletal hierarchies, we apply extensive structural perturbations. With probability $p=0.5$, we randomly delete up to 50\% of joints (and their associated mesh vertices) or remove entire subtrees. During VAE training, we also reconnect up to 30\% of joints to new parents ($p=0.5$), merging their skinning weights to simulate topology edits.}
\item \cyp{\textbf{Geometric Perturbation:} We apply non-uniform scaling ($p=0.5$), global rotations around principal axes ($p=0.2$, max $15^\circ$), and random pose perturbations ($p=0.5$, max 30° rotation). Additionally, we inject Gaussian noise into joint coordinates ($\sigma=10^{-2}$) and vertex positions ($\sigma=10^{-3}$) to enforce resilience against noisy inputs.}
\end{enumerate}

\subsubsection{\cyp{Model Architecture and Training}}

\added{Our framework cleanly decouples \emph{representation learning} from \emph{generation}. Stage~1 trains the \Representation FSQ-CVAE to compress the skinning weights of each individual bone into $T_D$ discrete tokens, while Stage~2 trains the \Framework autoregressive model to generate the full rig (the skeleton followed by the \Representation of every bone) as a single token sequence, which is then refined by GRPO post-training. Importantly, we reuse only the \emph{network structure} of the 3DShape2Vecset and Qwen3-0.6B designs and train both models entirely from scratch; no pretrained weights are involved. We consolidate the architectural and training configurations of both stages in Table~\ref{tab:model_config} and elaborate on each component below.}

\begin{table}[t]
\centering
\caption{\added{\textbf{Two-Stage Model and Training Configurations.} We summarize the architecture and training recipe of the \Representation FSQ-CVAE (Stage~1) and the \Framework autoregressive model (Stage~2). For both stages we adopt only the \emph{network structure} of the referenced designs and train from scratch. Stage~2 is further refined via GRPO post-training ($800$ steps, learning rate $10^{-6}$, group size $G\!=\!24$, $\epsilon\!=\!0.2$, $\beta\!=\!0.1$; see Sec.~\ref{sec:GRPO}).}}
\label{tab:model_config}
\resizebox{\columnwidth}{!}{
\added{
\begin{tabular}{l cc}
\toprule
 & \textbf{Stage 1: \Representation} & \textbf{Stage 2: \Framework} \\
 & \textbf{(FSQ-CVAE)} & \textbf{(Autoregressive)} \\
\midrule
\multicolumn{3}{l}{\textit{Architecture (structure only, trained from scratch)}} \\
\quad Network            & 3DShape2Vecset       & Qwen3-0.6B \\
\quad Parameters         & $110$M               & $\sim\!0.6$B \\
\quad Layers             & $2$ / $10$ (enc./dec.) & $0/28$ (decoder-only) \\
\quad Hidden width        & $768$                & $1024$ \\
\midrule
\multicolumn{3}{l}{\textit{Tokenization}} \\
\quad FSQ levels         & $[8,8,8,5,5,5]$      & $[8, 8, 8, 5, 5, 5]$ \\
\quad Codebook size       & $64{,}000$           & $64{,}000$ \\
\quad Shape tokens $T_W$  & $384$                & $384$ \\
\quad SkinTokens $T_D$    & $32$ (max)           & $8$ \\
\midrule
\multicolumn{3}{l}{\textit{Optimization}} \\
\quad Optimizer          & Muon\,+\,AdamW       & Muon\,+\,AdamW \\
\quad LR (Muon / AdamW)   & $2{\times}10^{-3}$ / $4{\times}10^{-4}$ & $2{\times}10^{-3}$ / $1{\times}10^{-4}$ \\
\quad Weight decay        & $0.01$               & $0.001$ \\
\quad Effective batch     & $320$                & $160$ \\
\quad Steps               & $400$k                & $300$k \\
\bottomrule
\end{tabular}
}
}
\end{table}

\paragraph{{\Representation (FSQ-CVAE)}}
We adopt the architecture of 3DShape 2Vecset~\cite{zhang20233dshape2vecset} ($110$M parameters), reusing only its network structure rather than any pretrained weights, with an asymmetric design (2 encoder layers, 10 decoder layers) to accelerate subsequent training, and PMPE encoding~\cite{bhat2025cube}. The FSQ codebook is configured with level sizes $[8,8,8,5,5,5]$, yielding a total vocabulary of $64,000$. During training, we used at most $T_D=32$ skin tokens and $T_W=384$ shape tokens as auxiliary conditions.

\paragraph{{\Framework (Autoregressive Model)}}
We adopt the Qwen3-0.6B architecture~\cite{qwen3}, utilizing Grouped Query Attention (GQA)~\cite{ainslie2023gqa} and Rotary Position Embeddings (RoPE)~\cite{su2024roformer} for efficient sequence modeling.

\paragraph{{Training}}
Both stages employ a hybrid optimizer strategy for efficiency: Muon~\cite{liu2025muon} is used for all attention-related layers, while AdamW~\cite{loshchilov2017decoupled} handles remaining parameters. The FSQ-CVAE is trained for $400$k iterations (batch size $320$), and \Framework for $300$k iterations (batch size $160$). \added{All models are trained on a single node with $8\times$ NVIDIA A800 (80\,GB) GPUs: the FSQ-CVAE (Stage~1) takes about $3$ days, the autoregressive \Framework (Stage~2) about $4$ days, and the GRPO refinement (Sec.~\ref{sec:GRPO}) about $3$ days, for a total of approximately $1{,}920$ GPU-hours.}

\subsubsection{RL Refinement (GRPO)}

For the post-training stage, we curate a smaller, high-complexity dataset of AI-generated meshes. We train for $800$ steps (learning rate $10^{-6}$ ) with a group size of $G=24$, clip ratio $\epsilon=0.2$, and KL penalty $\beta=0.1$. The reward weights are set to $w_{vj}=5$ and $w_{vk}=w_{sc}=w_{mo}=1$ (see Section~\ref{sec:GRPO}), emphasizing volumetric joint coverage while maintaining geometric validity.

\subsection{\cyp{Analysis of the \Representation Representation}}

\subsubsection{\cyp{Metrics and Reconstruction Fidelity}}
To substantiate our hypothesis that skinning weights exhibit intrinsic compressibility, we evaluate the reconstruction quality of the FSQ-CVAE across various codebook configurations. While Mean Absolute Error (MAE) is commonly used~\cite{liu2025riganything, xu2020rignet, song2025magicarticulate}, we argue it is insufficient for assessing skinning quality; due to the extreme sparsity of the matrix, a model can achieve low MAE by simply predicting near-zero values everywhere, failing to capture the critical active deformation zones. Therefore, we additionally employ \textit{Intersection over Union (IoU)}, a metric standard in semantic segmentation, to measure the structural overlap of the predicted influence regions. We categorize weights as active if they exceed a threshold $\varepsilon=10^{-2}$. For a ground truth weight $w_{i,j}$ and prediction $\hat{w}_{i,j}$, the IoU is defined as:

\[
\mathrm{IoU} =
\frac{
\sum_{i,j} [\,w_{i,j} > \varepsilon\, \land \,\hat{w}_{i,j} > \varepsilon\,]
}{
\sum_{i,j} [\,w_{i,j} > \varepsilon\, \lor \,\hat{w}_{i,j} > \varepsilon\,]
}.
\]
As illustrated in Figure~\ref{fig:recon_vae}, our model maintains high reconstruction fidelity even with a highly compact representation. The IoU scores remain robust as the number of tokens ($T_D$) decreases, demonstrating that the relevant information is effectively concentrated in the top few tokens. \added{Given that MAE alone can be dominated by the abundant zero entries, we recommend reporting IoU as a standard metric alongside MAE for skinning evaluation in future work.}

\subsubsection{\cyp{Compression Efficiency}}
\cyp{Table~\ref{tab:codebook} quantifies the compression performance. We compare different FSQ level configurations, specifically targeting codebook sizes of $15,360$ and $64,000$. Our selected configuration ($C = [8,8,8,5,5,5]$, size $64,000$) achieves a $183.74\times$ compression ratio compared to the raw FP16 representation, while maintaining a codebook utilization of \revised{$97.3\%$}. This confirms that the high-dimensional skinning matrix can be effectively reduced to a sequence of discrete \Representation (totaling just $64$ bytes per bone for $T_D = 32$) without significant information loss.}

\subsubsection{\added{Choice of Tokenizer: FSQ vs. VQ vs. Continuous}}
\added{A natural question is whether the Finite Scalar Quantization (FSQ) used in our \Representation is preferable to a learnable VQ-VAE codebook or a fully continuous CVAE. We compare all three under an identical backbone and token budget in Table~\ref{tab:tokenizer_compare}.}

\begin{table}[t]
\centering
\caption{\added{\textbf{Comparison of Tokenizer Choices.} We compare FSQ (ours), a learnable VQ-VAE codebook, and a continuous-latent CVAE under an identical backbone and token budget. Both FSQ and VQ are evaluated with the same codebook size of $64{,}000$, with all other settings kept consistent with the configuration described above. The $L_1$ error follows the same skinning reconstruction metric as Figure~\ref{fig:recon_vae} (in units of $10^{-2}$).}}
\label{tab:tokenizer_compare}
\resizebox{\columnwidth}{!}{
\added{
\begin{tabular}{lccc}
\toprule
Metric & FSQ (Ours) & VQ & CVAE (cont.) \\
\midrule
\multicolumn{4}{l}{\textit{Reconstruction fidelity}} \\
\quad IoU@$10^{-2}\uparrow$            & $88.0\%$   & $87.6\%$   & $94.3\%$ \\
\quad $L_1$ error ($\times 10^{-2}$)$\downarrow$   & $1.00$   & $1.03$   & $0.45$ \\
\midrule
\multicolumn{4}{l}{\textit{Code usage \& token repetition}} \\
\quad Codebook utilization$\uparrow$   & $97.29\%$ & $65.90\%$ & --- \\
\quad Token repetition rate @32$\downarrow$        & $2.29\%$  & $8.92\%$  & --- \\
\quad Samples with $\geq\!1$ consec. repeat$\downarrow$ & $2.07\%$ & $20.22\%$ & --- \\
\bottomrule
\end{tabular}
}
}
\end{table}

\added{As summarized in Table~\ref{tab:tokenizer_compare}, the continuous CVAE yields the best reconstruction but is incompatible with our discrete autoregressive formulation; the residual error introduced by discretization is discussed in Sec.~\ref{sec:limitations}. Among discrete tokenizers, FSQ and VQ are comparable in reconstruction, yet VQ exhibits markedly lower codebook utilization and tends to produce repeated tokens when encoding multiple tokens. This repetition propagates to the autoregressive sampling stage and necessitates a large repetition penalty, whereas FSQ avoids it by construction, motivating its use in \Representation.}

\subsubsection{\cyp{Latent Space Semantics}}
\cyp{To understand \textit{what} the encoder learns, we visualize the continuous latent space ($L_W$) prior to quantization. Figure~\ref{fig:t-sne} presents a t-SNE~\cite{maaten2008visualizing} projection of skinning latents from $300$ instances in the VRoid dataset, colored by their corresponding bone identifiers (mapped to the Mixamo template). We observe distinct, well-separated clusters corresponding to specific anatomical parts (e.g., Head, Hips, LeftLeg), despite the significant geometric variation across the source meshes. This indicates that \Representation capture a \textit{semantic structural prior}, e.g., learning an abstract representation of ``what a leg's skinning looks like'', rather than merely memorizing vertex indices. This learned invariance is key to the model's ability to cross-compress and generalize to diverse characters.}

\subsection{Comparison}

\begin{table*}[t]
    \centering
    \renewcommand{\tabcolsep}{5.5mm}
    \begin{tabular}{l|ccc|ccc}
    \hline
        & \multicolumn{3}{c|}{ModelsResource} & \multicolumn{3}{c}{Articulation 2.0}\\
    \cline{2-7}
        & J2J$\downarrow$ & J2B$\downarrow$ & B2B$\downarrow$ 
        & J2J$\downarrow$ & J2B$\downarrow$ & B2B$\downarrow$ \\
    \hline
    RigNet~\cite{xu2020rignet}                   & 3.901 & 2.412 & 2.213 & 7.376 & 5.841 & 4.802 \\
    MagicArticulate~\cite{song2025magicarticulate}          & 3.024 & 2.260 & 1.915 & 4.003 & 3.026 & 2.586 \\
    Puppeteer~\cite{song2025puppeteer}                & 3.841 & 2.881 & 2.475 & 3.033 & 2.300 & 1.923 \\
    UniRig~\cite{zhang2025unirig}                   & 3.390 & 2.592 & 1.890 & 3.115 & 2.211 & 1.926 \\
    \hline
        \Framework(4 skin tokens)             & 2.857 & 2.025 & 1.568 & 2.515 & 1.694 & 1.469 \\
    \Framework(6 skin tokens)             & \textbf{2.838} & 2.149 & 1.656 & 2.541 & 1.864 & 1.582 \\
    \Framework(4 skin tokens, w/ GRPO)       & 2.893 & \textbf{2.012} & \textbf{1.547} & \textbf{2.485} & \textbf{1.599} & \textbf{1.463} \\
    \Framework(6 skin tokens, w/ GRPO)       & 2.894 & 2.063 & 1.566 & 2.521 & 1.638 & 1.500 \\
    \hline
    \end{tabular}
    \caption{\textbf{Quantitative Comparison of Skeletal Generation.} We evaluate skeletal structure accuracy using Chamfer Distance metrics: Joint-to-Joint (J2J), Joint-to-Bone (J2B), and Bone-to-Bone (B2B) on the ModelsResource~\cite{Models-Resource} and Articulation 2.0~\cite{song2025magicarticulate} datasets. Lower values ($\downarrow$) indicate better performance. \Framework consistently outperforms state-of-the-art baselines across all metrics, demonstrating superior fidelity in joint placement and bone connectivity.}
    \label{tab:skeleton_metrics}
\end{table*}

\begin{table*}[t]
    \centering
    \renewcommand{\tabcolsep}{0.6mm}
    \begin{tabular}{l|ccccc|ccccc}
    \hline
        & \multicolumn{5}{c|}{ModelsResource~\cite{xu2020rignet}} & \multicolumn{5}{c}{Articulation 2.0~\cite{song2025magicarticulate}}\\
    \cline{2-11}
        & Skin L1$\downarrow$ & L1 Var.$\downarrow$ & Precision$\uparrow$ & Recall$\uparrow$ & Motion$\downarrow$
        & Skin L1$\downarrow$ & L1 Var.$\downarrow$ & Precision$\uparrow$ & Recall$\uparrow$ & Motion$\downarrow$\\
    \hline
    RigNet~\cite{xu2020rignet}                   & 0.0573 & 0.0464 & 62.4 & 59.9 & 0.0789 & 0.0431 & 0.0395 & 67.8 & 54.6 & 0.0915\\
    Puppeteer~\cite{song2025puppeteer}                & 0.0321 & 0.0173 & 64.4 & 87.2 & 0.0279 & 0.0278 & 0.0144 & 76.7 & 75.1 & 0.0314\\
    UniRig~\cite{zhang2025unirig}                   & 0.0381 & 0.0212 & 65.8 & 86.7 & 0.0312 & 0.0297 & 0.0165 & 72.6 & 73.5 & 0.0419\\
    \hline
        \Framework (4 skin tokens)             & 0.0168 & 0.0072 & 78.9 & 88.1 & 0.0184 & 0.0178 & 0.0077 & 78.1 & 87.7 & 0.0236\\
    \Framework (6 skin tokens)             & 0.0166 & 0.0069 & 79.1 & \textbf{89.1} & 0.0176 & 0.0153 & 0.0061 & 78.8 & 88.8 & 0.0228\\
    \Framework(4 skin tokens, w/ GRPO)       & 0.0169 & 0.0071 & 78.9 & 88.3 & 0.0166 & 0.0174 & 0.0069 & 78.1 & 87.7 & 0.0214\\
    \Framework (6 skin tokens, w/ GRPO)       & \textbf{0.0163} & \textbf{0.0068} & \textbf{79.2} & \textbf{89.1} & \textbf{0.0158} & \textbf{0.0150} & \textbf{0.0058} & \textbf{79.0} & \textbf{89.2} & \textbf{0.0209} \\

    \hline
    \end{tabular}
    \caption{\textbf{Quantitative Evaluation of Skinning Prediction.} We compare skinning fidelity using L1 Error, Precision/Recall, and Motion Loss. A threshold of $\varepsilon=10^{-2}$ is applied to filter negligible weights. \Framework achieves superior performance across all categories, demonstrating significantly lower reconstruction error and higher precision/recall than baseline regression approaches.}
    \label{tab:skin_metrics}
\end{table*}

\begin{figure}[h!]
    \centering
    \includegraphics[width=0.5\textwidth]{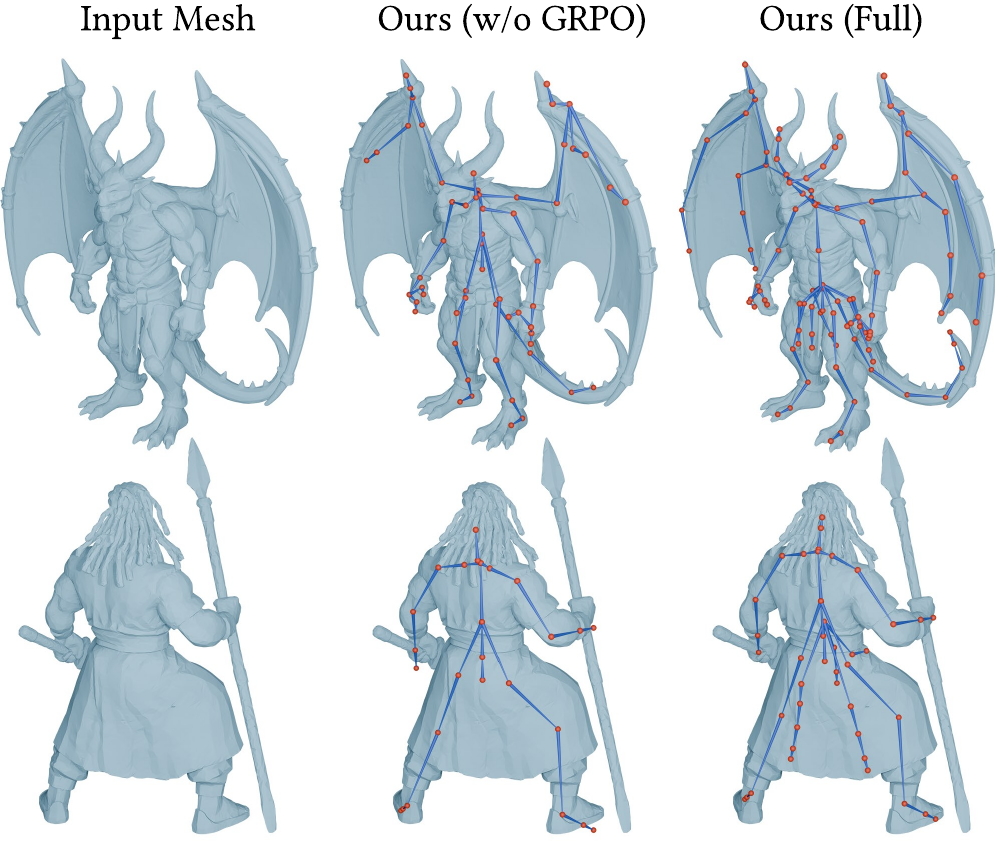}
    \caption{\textbf{Impact of GRPO on Skeletal Topology.} While supervised training alone often misses non-standard anatomy, the GRPO-refined model effectively synthesizes auxiliary bones for secondary structures, such as tails, horns, wings, and clothing accessories.}
    \label{fig:compare_grpo_skeleton}
\end{figure}

\begin{figure}[h!]
    \centering
    \includegraphics[width=0.5\textwidth]{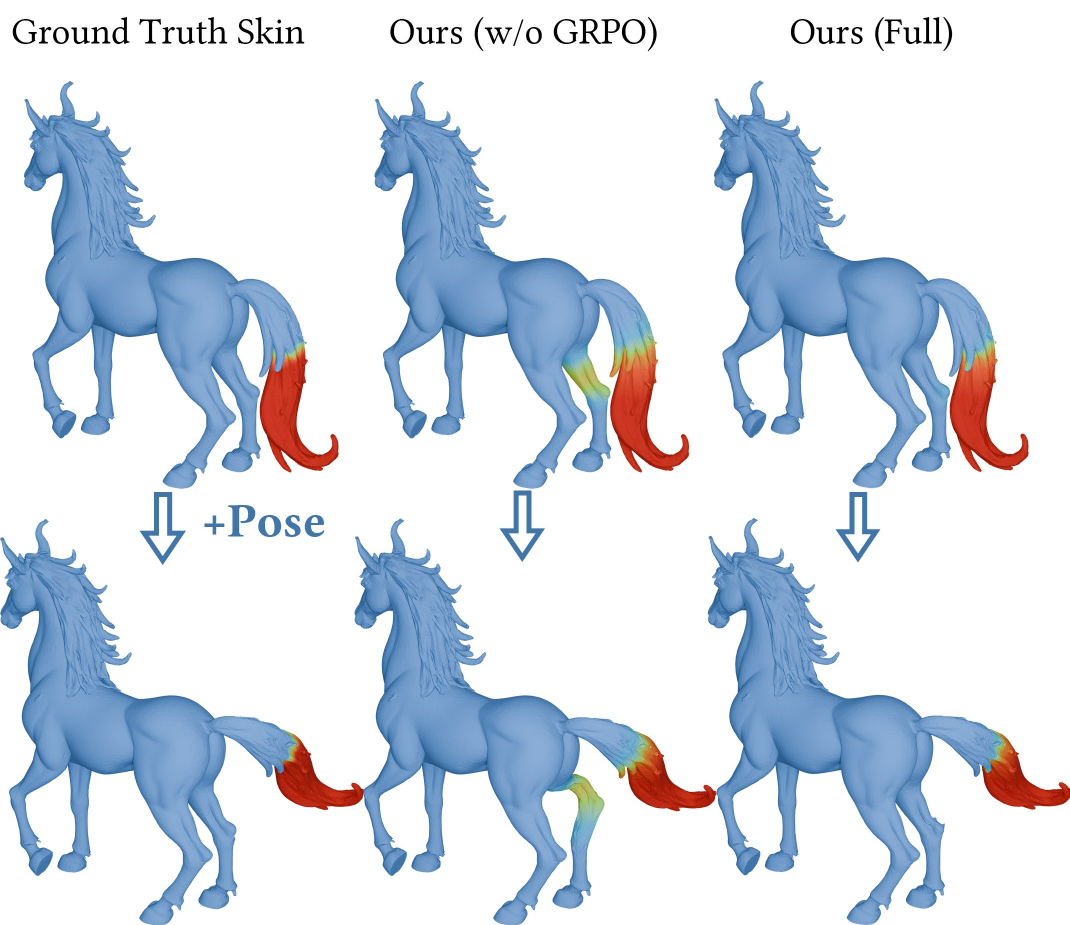}
    \caption{\textbf{Impact of GRPO on Skinning Precision.} In addition to topology, the reinforcement learning stage tightens skinning predictions. The model produces precise, locally distinct influence maps that minimize artifacts, yielding valid rigs suitable for high-quality animation.}
    \label{fig:compare_grpo_skin}
\end{figure}

We evaluate our framework against four leading learning-based rigging approaches: RigNet~\cite{xu2020rignet}, MagicArticulate~\cite{song2025magicarticulate}, UniRig~\cite{zhang2025unirig}, and Puppeteer~\cite{song2025puppeteer}. Evaluations are conducted on the ModelsResource~\cite{Models-Resource,xu2020rignet} and Articulation 2.0~\cite{song2025magicarticulate} test sets. All skeletons are normalized to the $[-1, 1]^3$ cube.

\subsubsection{Skeletal Generation Quality} 
\cyp{We assess skeletal structure using Chamfer Distance metrics: Joint-to-Joint (J2J), Joint-to-Bone (J2B), and Bone-to-Bone (B2B), as introduced in RigNet~\cite{xu2020rignet}.} As summarized in Table~\ref{tab:skeleton_metrics}, \Framework consistently outperforms all baselines across both datasets. \cyp{Notably, we achieve both the lowest J2J and B2B errors, indicating that our generated skeletons are not only closer to ground truth joints but also maintain superior topological alignment.}

These quantitative gains are reflected in the qualitative comparisons shown in Figure~\ref{fig:compare_rigging}. 
Baseline methods exhibit distinct structural weaknesses: RigNet~\cite{xu2020rignet} frequently generates incomplete skeletons, often missing terminal chains due to the limitations of its MST-based connectivity inference; conversely, UniRig~\cite{zhang2025unirig} tends to over-segment the mesh, producing an excessive number of joints with irregular topology that is difficult to animate. While newer autoregressive models like Puppeteer~\cite{song2025puppeteer} and MagicArticulate~\cite{song2025magicarticulate} improve upon this, they struggle to preserve fine-grained semantic details, often failing to capture anatomical features such as ears or horns on non-humanoid characters. In contrast, \Framework yields structurally coherent and semantically faithful skeletons, effectively balancing geometric coverage with topological simplicity.

\subsubsection{Skinning Prediction Fidelity}
\cyp{The most significant performance gains are observed in skinning prediction, validating the efficacy of the \Representation representation. We employ five complementary metrics: Precision and Recall (quantifying the accuracy of joint influence regions), Motion Loss (measuring deformation fidelity under LBS), $L1$ Error, and $L1$ Variance (reflecting the consistency of errors). Measurements are computed with a weight threshold $\varepsilon=10^{-2}$.} As reported in Table~\ref{tab:skin_metrics}, our method achieves state-of-the-art performance across all skinning metrics. 
\cyp{We observe a substantial reduction in $L1$ Error compared to RigNet~\cite{xu2020rignet} ($0.0163$ vs. $0.0573$ on ModelsResource), corresponding to the 98\%–133\% improvement highlighted in our findings, confirming that our discrete token prediction avoids the noise and ``mean-seeking'' behavior typical of continuous regression tasks. Also, the low $L1$ Variance indicates that \Framework produces consistently high-quality weights across different vertices, avoiding the localized failures common in baseline methods. Furthermore, the superior Motion Loss scores demonstrate that our predicted weights result in lower distortion when applied to actual deformations.}

\cyp{Qualitatively, these improvements translate to cleaner, more distinct segmentation. Figure~\ref{fig:compare_skin} illustrates that while baselines like UniRig~\cite{zhang2025unirig} and MagicArticulate~\cite{song2025magicarticulate} often produce bleeding artifacts, i.e., where skinning weights spill onto disconnected mesh components. Our FSQ-CVAE decoder enforces strict locality, resulting in artifact-free weight maps. This precision is particularly evident in complex articulations; for example, in the third row of Figure~\ref{fig:compare_skin}, \Framework is the only method that accurately preserves the fine-grained spatial differentiation of finger joints, whereas baseline methods tend to predict over-smoothed or bleeding weights that degrade animation fidelity. Furthermore, our superior Motion Loss scores indicate that these weights are not just statically accurate but produce lower distortion when applied to actual deformations.}

\subsubsection{\added{Inference Efficiency}}
\added{As \Framework generates the rig autoregressively, its runtime grows with the joint count $J$ and per-bone token count $T_D$. Table~\ref{tab:inference_time} reports the skeleton and skinning time against the baselines (averaged over the test set on a single A800). \Framework completes a full rig in $\sim\!15$s, a favorable accuracy--latency trade-off for an end-to-end pipeline.}

\begin{table}[h!]
\centering
\caption{\added{\textbf{Inference Time Comparison.} We report the average wall-clock time for skeleton and skinning prediction. RigNet's runtime is highly sensitive to the predicted joint count. Puppeteer's skinning stage is dominated by a marching-cubes volumetric step. \Framework achieves a full rig in $\sim\!15$s, with skinning decoded directly from \Representation.}}
\label{tab:inference_time}
\resizebox{\columnwidth}{!}{
\added{
\begin{tabular}{lccc}
\toprule
Method & Skeleton & Skinning & Total \\
\midrule
RigNet~\cite{xu2020rignet}            & \multicolumn{3}{c}{$1$s\,--\,$20$min (joint-count dependent)} \\
UniRig~\cite{zhang2025unirig}         & $\sim\!4$s & $\sim\!1$s & $\sim\!5$s \\
Puppeteer~\cite{song2025puppeteer}    & $\sim\!4$s & $\sim\!30$s (marching cubes) & $\sim\!34$s \\
\midrule
\Framework (Ours)                     & $\sim\!4$s & $10$--$12$s & $\sim\!15$s \\
\bottomrule
\end{tabular}
}
}
\end{table}

\begin{table*}[h!]
    \centering
    \begin{tabular}{c|ccc|ccc|ccc}
    \hline
        & \multicolumn{3}{c|}{VRoid Hub} & \multicolumn{3}{c|}{ModelResource} & \multicolumn{3}{c}{Articulation 2.0}\\
    \cline{2-10}
        & L1 Error$\downarrow$ & IoU$\uparrow$ & Mask$^*\uparrow$ & L1 Error$\downarrow$ & IoU$\uparrow$ & Mask$^*\uparrow$ & L1 Error$\downarrow$ & IoU$\uparrow$ & Mask$^*\uparrow$ \\
    \hline
    \Framework & $\bf{5.41 \times 10^{-3}}$ & \bf{87.1 \%} & \bf{92.5 \%} & $1.38 \times 10^{-2}$ & \bf{91.1 \%} & \bf{83.9 \%} & $1.40 \times 10^{-2}$ & \bf{84.0 \%} & \bf{82.2 \%} \\
    - w/o Dice loss & $5.98 \times 10^{-3}$ & 82.2 \% & 91.6 \% & $\bf{1.35 \times 10^{-2}}$ & 88.2 \% & 82.7 \% & $\bf{1.36 \times 10^{-2}}$ & 80.9 \% & 80.8 \%  \\
    \hline
    \end{tabular}
    \caption{\textbf{Ablation Study on Loss Function Design.} We evaluate the impact of the Dice Loss term on reconstruction quality, using a baseline configuration with codebook size $C=[8,8,8,5,6]=15,360$ and $T_D = 32$ skin tokens. The results show that removing Dice supervision significantly degrades IoU performance across all datasets. $^*$Mask Accuracy quantifies the proportion of samples where the predicted non-zero support fully covers the ground truth active region (tolerance $\varepsilon=10^{-2}$), highlighting the loss's role in preventing under-segmentation.}
    \label{tab:dice_augmentation}
\end{table*}

\begin{table*}[t!]
    \centering
    \renewcommand{\tabcolsep}{5.5mm}
    \begin{tabular}{l|ccc|ccc}
    \hline
        & \multicolumn{3}{c|}{ModelsResource} & \multicolumn{3}{c}{Articulation 2.0}\\
    \cline{2-7}
        & J2J$\downarrow$ & J2B$\downarrow$ & B2B$\downarrow$ 
        & J2J$\downarrow$ & J2B$\downarrow$ & B2B$\downarrow$ \\
    \hline
    \Framework                        & \textbf{2.857} & \textbf{2.025} & \textbf{1.568} & \textbf{2.515} & \textbf{1.694} & \textbf{1.469} \\
    - w/o non-uniform scaling       & 3.030 & 2.171 & 1.689 & 2.679 & 1.844 & 1.618 \\
    - w/o sub-tree dropping         & 3.019 & 2.179 & 1.704 & 2.648 & 1.806 & 1.582 \\
    - w/o joints deleting & 3.077 & 2.220 & 1.742 & 2.818 & 1.977 & 1.713 \\
    \hline
    \end{tabular}
    \caption{\textbf{Ablation Study on Data Augmentation Strategies.} We analyze the contribution of each robustness-oriented augmentation module to skeletal prediction accuracy (J2J, J2B, B2B). Experiments are conducted using a consistent lightweight configuration ($T_D=4$). The results indicate that removing any augmentation component (particularly random joint deletion) leads to higher error rates, confirming that simulating imperfections is essential for achieving robust generalization.}
    \label{tab:skeleton_augmentation}
\end{table*}

\subsection{Ablation Study}

\cyp{We conduct a series of ablation studies to validate the critical components of our framework, specifically analyzing the impact of the loss function design, the reinforcement learning stage, and our data augmentation strategy.}

\subsubsection{Necessity of Dice Loss}

As discussed in Section~\ref{sec:loss}, the skinning weight matrix is characterized by extreme sparsity, leading to a severe class imbalance between active and inactive weights. We hypothesized that the Dice Loss~\cite{sudre2017generalised} is essential to mitigate optimization difficulties arising from this imbalance. To evaluate this, we trained a variant of our model using only Binary Cross Entropy (BCE) and Mean Squared Error (MSE), setting $\lambda_\text{Dice}=0$. The quantitative results, presented in Table~\ref{tab:dice_augmentation}, confirm our hypothesis. Removing the Dice Loss leads to a significant degradation in reconstruction accuracy, with IoU scores dropping from $87.1\%$ to $82.2\%$ on the VRoid dataset. \cyp{Empirically, we observed that without Dice supervision, the model tends to under-predict active regions, struggling to distinguish subtle weight gradients from zero-background noise. The inclusion of Dice Loss is thus critical for achieving stable convergence and high-fidelity sparse reconstruction.}

\subsubsection{Effectiveness of GRPO Training}
\cyp{A key motivation for our unified framework is the ability to leverage reinforcement learning to generalize beyond the supervised training distribution.} We assessed the impact of our GRPO-based post-training on both standard metrics and out-of-distribution (OOD) cases. As indicated in Table~\ref{tab:skeleton_metrics} and Table~\ref{tab:skin_metrics}, the model fine-tuned with GRPO maintains or improves performance on standard benchmarks.

\cyp{The true value of GRPO, however, lies in its extrapolation capability.} Figure~\ref{fig:compare_grpo_skeleton} and Figure~\ref{fig:compare_grpo_skin} visualize the qualitative gains on complex, in-the-wild meshes that differ significantly from the training data. In the base model, we occasionally observed failure cases where auxiliary structures such as demonic wings, capes, or tails were ignored or received ambiguous skinning. The GRPO-trained model, guided by the volumetric coverage and bone-mesh containment rewards, successfully synthesizes coherent bone hierarchies for these challenging structures. For instance, it accurately places bones within the thin geometry of wings and differentiates the skinning of closely interacting regions, such as horse tails and hind legs. These results suggest that our reward-based refinement effectively injects geometric reasoning into the generation process, enhancing robustness where supervised signals are scarce.

\begin{table*}[t]
    \centering
    \added{
    \renewcommand{\tabcolsep}{4mm}
    \begin{tabular}{l|ccc|ccc}
    \hline
        & \multicolumn{3}{c|}{Skeleton (Chamfer Distance)} & \multicolumn{3}{c}{Skinning} \\
    \cline{2-7}
        & J2J$\downarrow$ & J2B$\downarrow$ & B2B$\downarrow$ & $L_1\downarrow$
        & IoU@$10^{-3}\uparrow$ & IoU@$10^{-1}\uparrow$ \\
    \hline
    \Framework\ (w/o GRPO) & $2.459$ & $1.759$ & $1.412$ & $9.84\!\times\!10^{-3}$ & $0.562$ & $0.710$ \\
    \Framework\ (w/ GRPO)        & ${2.388}\gain{2.9}$ & ${1.717}\gain{2.4}$ & ${1.394}\gain{1.3}$ & ${9.57\!\times\!10^{-3}}\gain{2.7}$ & ${0.571}\gain{1.6}$ & ${0.714}\gain{0.5}$ \\
    \hline
    \end{tabular}
    }
    \caption{\added{\textbf{Held-out OOD Evaluation of GRPO Post-training (Truebones Zoo).} We report skeletal Chamfer distances (J2J/J2B/B2B), the skinning $L_1$ error, and the skinning IoU at thresholds $\varepsilon\in\{10^{-3},10^{-1}\}$ on the Truebones~Zoo~\cite{truebones} dataset, which is entirely held out from training and orthogonal to our training corpus. GRPO post-training improves \emph{all} metrics; the relative improvement over the base model is annotated in red (\textnormal{\color{red}(+x\%)}), providing direct quantitative evidence that GRPO enhances out-of-distribution extrapolation rather than overfitting the training distribution.}}
    \label{tab:grpo_ood}
\end{table*}

\added{\noindent\textbf{Held-out OOD Benchmark.} To rigorously quantify this extrapolation capability beyond qualitative cases, we further evaluate on a fully held-out benchmark that is \emph{orthogonal} to our training corpus. Inspired by MoCapAnything~\cite{gong2026mocapanything}, we adopt the Truebones~Zoo~\cite{truebones} dataset, a large collection of artist-created animal rigs ($1{,}038$ motion sequences totaling $104{,}715$ frames, spanning a broad range of species and kinematic structures) that never appear during either the supervised or the GRPO stage. As reported in Table~\ref{tab:grpo_ood}, GRPO post-training improves \emph{every} metric on this unseen benchmark, lowering the skeletal Chamfer distances (J2J/J2B/B2B) and the skinning $L_1$ error while raising the IoU across all thresholds. These consistent held-out gains, annotated in red, substantiate that our reward-driven refinement enhances generalization to genuinely out-of-distribution structures rather than overfitting the training distribution.}

\subsubsection{\added{Deformation Smoothness: Isolating Tokenization from RL Refinement}}
\added{A compressed representation could attain a low reconstruction error yet still introduce non-smooth skinning. To rule this out and disentangle the smoothness contribution of tokenization from that of the RL stage, we evaluate the deformation-smoothness reward $R_{mo}$ (higher is better; Section~\ref{sec:GRPO}) at three stages: (i)~the FSQ tokenizer reconstruction alone, (ii)~the autoregressive model before GRPO, and (iii)~after GRPO. Scores are normalized against the artist-authored Ground Truth ($100\%$ reference) on the held-out Truebones~Zoo~\cite{truebones} and in-distribution Articulation~2.0~\cite{song2025magicarticulate} datasets (Table~\ref{tab:smoothness}).}

\begin{table}[t]
\centering
\caption{\added{\textbf{Isolating Tokenization from RL Refinement on Deformation Smoothness.} We report the deformation-smoothness reward $R_{mo}$ (higher is better; Section~\ref{sec:GRPO}); each cell lists the raw $R_{mo}$ together with its percentage relative to the Ground Truth skinning, shown in \textcolor{red}{red} parentheses. \emph{FSQ Reconstruction} evaluates the tokenizer alone (no autoregressive generation, no RL) at $T_D\!\in\!\{4,8\}$ tokens per bone; \emph{Autoregressive} evaluates the full generation ($T_D{=}4$) before and after GRPO. Truebones~Zoo~\cite{truebones} is fully held out (OOD), whereas Articulation~2.0~\cite{song2025magicarticulate} is in-distribution. Tokenization alone retains most of the Ground Truth smoothness (and more so with a larger token budget), while the larger drop from autoregressive sampling is substantially recovered by GRPO.}}
\label{tab:smoothness}
\setlength{\tabcolsep}{4pt}
\renewcommand{\arraystretch}{1.1}
\added{
\resizebox{\columnwidth}{!}{%
\begin{tabular}{lccccc}
\toprule
 & \multicolumn{2}{c}{FSQ Recon.} & \multicolumn{2}{c}{Autoregressive} & \\
\cmidrule(lr){2-3}\cmidrule(lr){4-5}
Dataset & $T_D{=}4$ & $T_D{=}8$ & w/o GRPO & w/ GRPO & GT \\
\midrule
Truebones Zoo & \makecell{$0.5748$\\{\scriptsize\textcolor{red}{($96.4\%$)}}} & \makecell{$0.5816$\\{\scriptsize\textcolor{red}{($97.6\%$)}}} & \makecell{$0.5473$\\{\scriptsize\textcolor{red}{($91.8\%$)}}} & \makecell{$0.5684$\\{\scriptsize\textcolor{red}{($95.3\%$)}}} & \makecell{$0.5962$\\{\scriptsize\textcolor{red}{($100\%$)}}} \\
Articulation 2.0 & \makecell{$0.2711$\\{\scriptsize\textcolor{red}{($87.3\%$)}}} & \makecell{$0.2872$\\{\scriptsize\textcolor{red}{($92.4\%$)}}} & \makecell{$0.2289$\\{\scriptsize\textcolor{red}{($73.7\%$)}}} & \makecell{$0.2505$\\{\scriptsize\textcolor{red}{($80.6\%$)}}} & \makecell{$0.3107$\\{\scriptsize\textcolor{red}{($100\%$)}}} \\
\bottomrule
\end{tabular}%
}
}
\end{table}

\added{Two conclusions follow. First, the FSQ reconstruction --- trained without any smoothness reward --- already preserves most of the Ground Truth smoothness ($96.4\%$/$87.3\%$ on Truebones~Zoo/Articulation~2.0 at $T_D{=}4$), improving with the token budget ($97.6\%$/$92.4\%$ at $T_D{=}8$); discretization thus incurs only a modest cost that shrinks with more tokens, confirming that a low $L_1$/IoU error does not conceal non-smooth artifacts. Second, the larger drop stems from autoregressive sampling ($91.8\%$/$73.7\%$ before GRPO), which GRPO substantially recovers ($95.3\%$/$80.6\%$)---as expected, since $R_{mo}$ is one of its objectives---compensating for this generation-induced gap rather than for tokenization loss.}

\subsubsection{Effectiveness of Data Augmentation}
Finally, we verify the importance of our robustness-oriented data augmentation pipeline. \cyp{We performed an ablation by selectively disabling specific augmentation modules, such as non-uniform scaling, sub-tree dropping, and random joint deletion, and retraining the autoregressive model. As reported in Table~\ref{tab:skeleton_augmentation}, removing any single module results in a consistent degradation of skeletal prediction accuracy across all datasets. For example, omitting non-uniform scaling increases the J2J error on ModelsResource from $2.857$ to $3.030$. The most significant drop occurs when removing joint deletion strategies, confirming that simulating topological imperfections during training is vital for handling the diverse and often noisy geometry found in production environments.}

\subsection{\added{Additional Qualitative Results}}
\begin{figure*}[htbp]
    \centering
    \includegraphics[width=\textwidth]{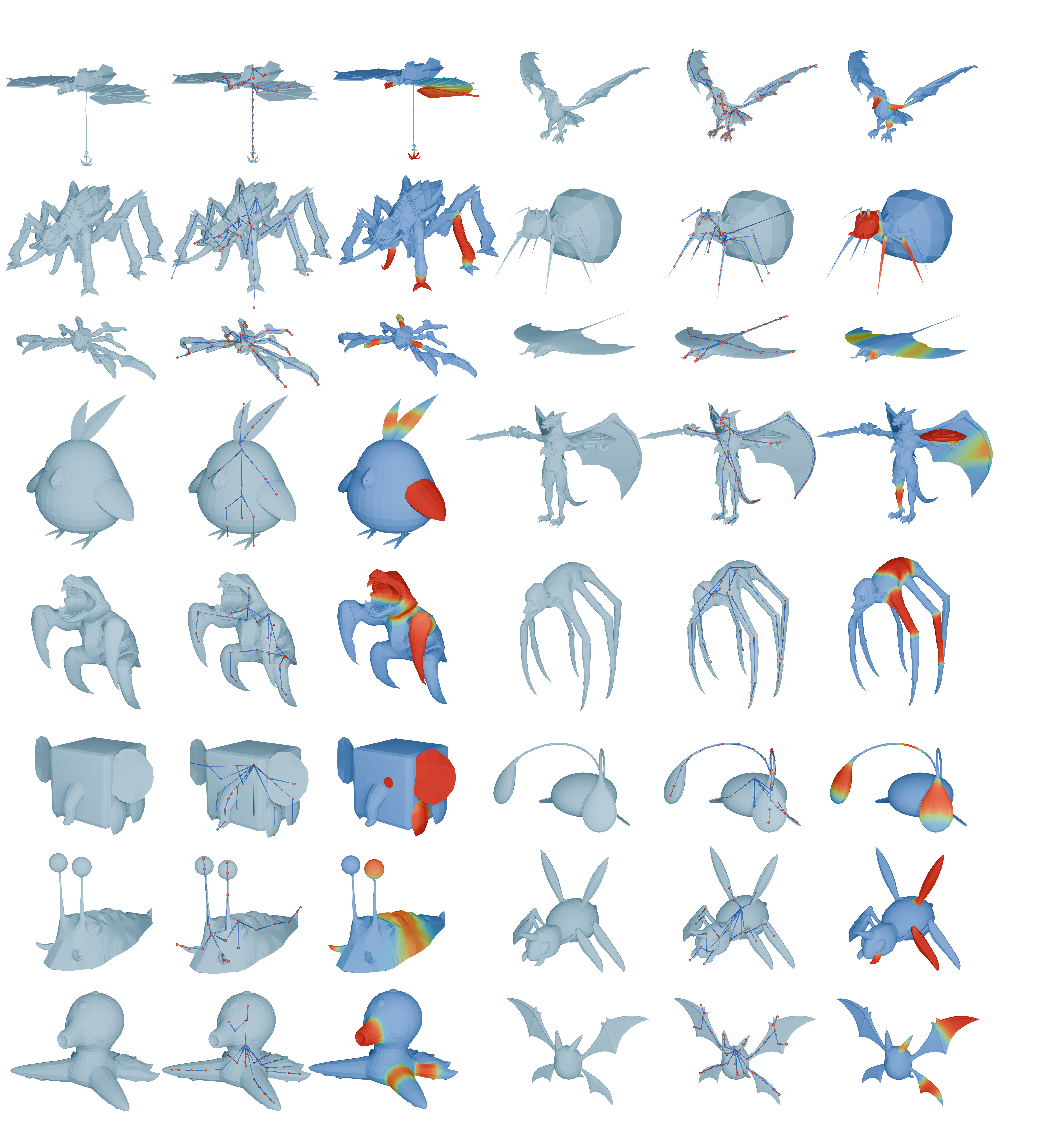}
    \caption{\added{\textbf{Additional Qualitative Results.} An extended gallery of rigs produced by \Framework across diverse characters, including birds, insects, dragons, and other non-humanoid creatures. Each triplet shows the input mesh, the generated skeleton, and the predicted skinning weights (heatmap), illustrating robust generalization to varied morphologies and topologies.}}
    \label{fig:more_results}
\end{figure*}

\added{To further demonstrate the generalization of \Framework beyond the comparisons above, Figure~\ref{fig:more_results} presents an extended gallery of rigging results spanning a broad range of character morphologies and skeletal topologies. For each asset, we show the input mesh, the generated skeleton, and the predicted skinning weights (visualized as a heatmap).}

\section{Conclusion}

In this work, we introduced \textit{\Framework}, an automated framework for skeletal rigging and skinning weight prediction that approaches the fidelity of professional artist workflows. Our core insight is that the longstanding bottleneck in automatic skinning is a representation problem. By proposing \textit{\Representation}, i.e., a compact, discrete representation learned via an FSQ-CVAE, we successfully transformed the ill-posed regression of sparse skinning matrices into a robust token prediction task. This design enabled us to train a unified autoregressive model that jointly synthesizes skeletal structures and skinning weights, capturing the intrinsic dependencies between articulation and deformation that prior decoupled methods ignore. Our experiments demonstrate that this unified approach, combined with a novel GRPO-based reinforcement learning stage, significantly outperforms state-of-the-art baselines. The integration of geometric and semantic reward functions proves particularly effective for generalization, allowing the model to generate coherent rigs for complex, out-of-distribution assets that defy standard supervised learning.

\begin{figure}[H]
    \centering
    \includegraphics[width=0.7\linewidth]{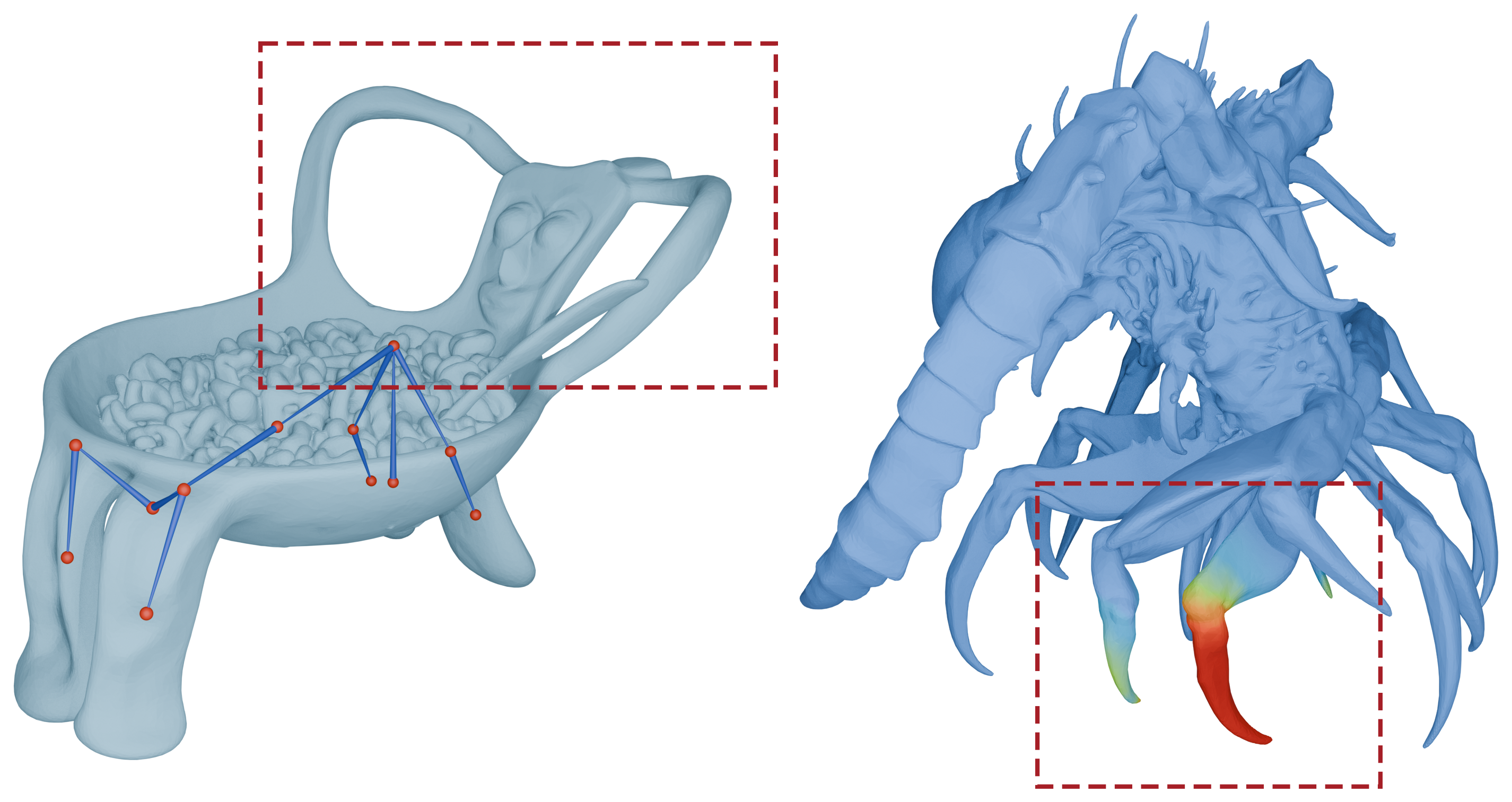}
    \caption{\added{\textbf{Failure Cases.} \textbf{Left:} on a thin, looped handle-like structure, the predicted skeleton deviates from the intended articulation (highlighted). \textbf{Right:} a slender claw exhibits a residual skinning spike, where a few vertices are erroneously bound to a distant bone, producing a localized high-weight artifact (highlighted). These failures mainly occur on out-of-distribution geometries with extremely thin or ambiguous topology.}}
    \label{fig:failure_cases}
\end{figure}

\noindent \phantomsection\label{sec:limitations}\textbf{Limitations and Discussions.}
Despite these advancements, several avenues for improvement remain. \added{We illustrate representative failure cases in Figure~\ref{fig:failure_cases}, which predominantly stem from extremely thin or topologically ambiguous geometry that is under-represented in training.} First, while our FSQ-CVAE offers high compression efficiency, a comparative analysis suggests a residual performance gap remains compared to continuous-latent VAEs in extremely challenging skinning scenarios. \revised{As quantified in Table~\ref{tab:tokenizer_compare}, discretization does incur a measurable reconstruction loss relative to a continuous CVAE, so our use of discrete tokens is a deliberate trade-off: it enables a \emph{unified} autoregressive model (skeletal parameters are inherently discrete) and fast sampling without iterative denoising. We expect continuous representations to eventually close this gap as mainstream generative paradigms mature, while the underlying idea of \emph{compressing} skinning weights should remain central regardless of whether the tokens are discrete or continuous.} Recent developments in continuous token representations~\cite{sikder2025transfusion, li2024autoregressive} may offer a path to bridge this gap, potentially enhancing predictive precision without sacrificing the benefits of sequence modeling. Second, our current framework generates rigs autonomously based on learned priors. However, professional production often requires adherence to specific topological standards. Extending our autoregressive model to accept user-specified topological templates or interactive guidance would be a valuable direction, effectively transforming \Framework from an automatic generator into a flexible, artist-directed co-pilot. Finally, while our RL stage improves geometric validity, future work could explore physics-based rewards to further ensure the dynamic plausibility of the generated deformations during animation.

\begin{figure*}[tp]
    \centering
    \includegraphics[width=0.9\textwidth]{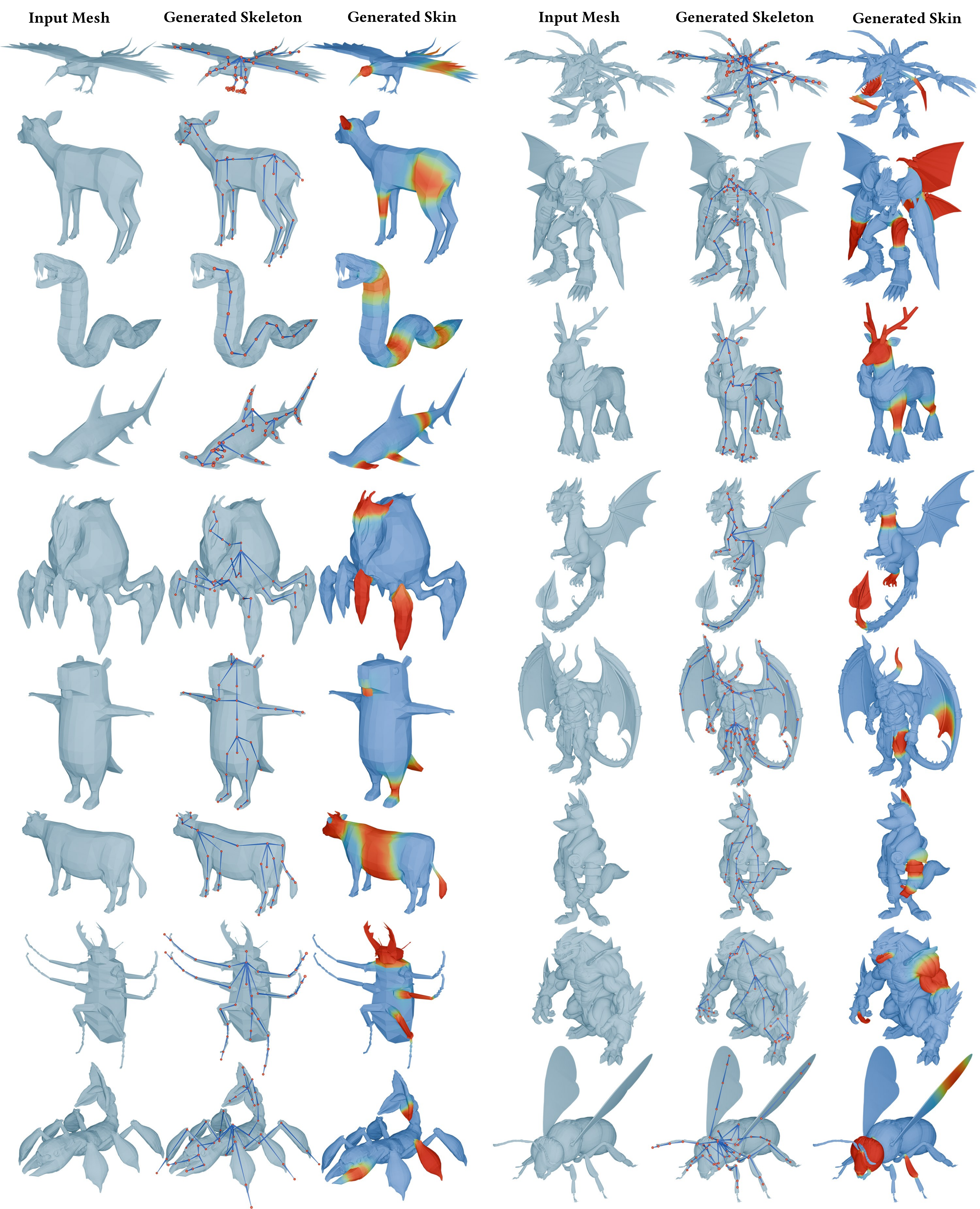}
    \caption{\textbf{Diverse Generation Results.} We demonstrate the generalization capacity of \Framework on a wide range of inputs, including unseen test-set samples and complex in-the-wild assets. The model robustly synthesizes fully articulated skeletons and accurate skinning weights.}
    \label{fig:our_results}
\end{figure*}

\bibliographystyle{ACM-Reference-Format}
\bibliography{ref}

\end{document}